\documentclass[prd,aps,twocolumn,nofootinbib,showpacs,superscriptaddress,amsmath,amssymb,aps]{revtex4-1}
\usepackage[bottom]{footmisc}
\usepackage{graphicx}% Include figure files
\usepackage{dcolumn}% Align table columns on decimal point
\usepackage{bm}% bold math
\usepackage{natbib}

\def\fun#1#2{\lower3.6pt\vvbox{\baselineskip0pt\lineskip.9pt
  \ialign{$\mathsurround=0pt#1\hfil##\hfil$\crcr#2\crcr\sim\crcr}}}
\def\simgt{\mathrel{\lower0.6ex\hbox{$\buildrel {\textstyle >}
 \over {\scriptstyle \sim}$}}}
\def\simlt{\mathrel{\lower0.6ex\hbox{$\buildrel {\textstyle <}
 \over {\scriptstyle \sim}$}}}

\def\bea{\begin{eqnarray}}
\def\eea{\end{eqnarray}}
\def\be{\begin{equation}}
\def\ee{\end{equation}}
\def\bes{\begin{split}}
\def\ees{\end{split}}
\def\ba{\begin{eqnarray}}
\def\ea{\end{eqnarray}}
\def\p{\partial}

\def\ruh{r_{\rm uh}}
\def\dd{{\rm d}}
\def\dt{{\rm d}t}
\def\dv{{\rm d}v}
\def\dr{{\rm d}r}
\def\dx{{\rm d}x}

\def\dz{{\rm d}z}

\def\ds{{\rm d}s}
\def\dth{{\rm d}\theta}
\def\dOm{{\rm d}\Omega}
\def\dvph{{\rm d}\varphi}

\def\dg{\delta g}
\def\m{{\rm m}}
\def\({\Big(}
\def\){\Big)}

\def\mO{{\mathcal O}}

\def\vv{{\rm v}}
\def\dv{{\rm dv}}

\def\mn{{\mu\nu}}

\def\tir{{\tilde{r}}}

\def\a{\alpha}
\def\b{\beta}
\def\c{\chi}
\def\r{\rho}
\def\s{\sigma}

\def\m{\mu}
\def\n{\nu}

\def\th{\theta}

\def\S{\Sigma}
\def\g{\gamma}
\def\l{\lambda}

\def\d{\delta}
\def\kh{{\rm kh}}

\newcommand{\um}{^{\mu}}

\newcommand{\lmn}{_{\mu\nu}}

\newcommand{\lab}{_{\alpha\beta}}

\newcommand\bw{\begin{widetext}}
\newcommand\ew{\end{widetext}}

\newcommand{\nn}{\nonumber}

\begin{document}

\title{Constraints on Ho\v{r}ava gravity from binary black hole observations}% Force line breaks with \\

\author{Oscar Ramos}
\affiliation{Institut d'Astrophysique de Paris, CNRS \& Sorbonne
 Universit\'es, UMR 7095, 98 bis bd Arago, 75014 Paris, France}
\affiliation{Institut Lagrange de Paris (ILP), Sorbonne Universit\'es, 98 bis bd Arago, 75014 Paris, France}
\author{Enrico Barausse}
\affiliation{Institut d'Astrophysique de Paris, CNRS \& Sorbonne
 Universit\'es, UMR 7095, 98 bis bd Arago, 75014 Paris, France}

\date{\today}% It is always \today, today,

\begin{abstract}
Ho\v rava gravity breaks Lorentz symmetry by introducing a preferred spacetime foliation, which is defined by a timelike dynamical scalar field, the khronon. 
The presence of this preferred foliation makes black hole solutions more complicated than in General Relativity, with the appearance of multiple distinct
event horizons: a matter horizon for light/matter fields; a spin-0 horizon for the scalar excitations of the khronon; a spin-2 horizon for tensorial gravitational waves; and even, at least in spherical symmetry, a universal horizon for instantaneously propagating modes appearing in the ultraviolet. We study how black hole solutions in Ho\v rava gravity change when the black hole is allowed to move with low velocity relative to the preferred foliation. These slowly moving solutions are 
a crucial ingredient to compute black hole ``sensitivities'' and predict gravitational wave emission (and in particular dipolar radiation) from the inspiral of binary
black hole systems. We find that for {\it generic} values of the theory's three dimensionless coupling constants, slowly moving black holes present curvature singularities 
at the universal horizon. Singularities at the spin-0 horizon also arise unless one
waives the requirement of asymptotic flatness at spatial infinity. Nevertheless, we have verified that 
at least in a one-dimensional subset of the (three-dimensional) parameter space of the theory's coupling constants, slowly moving black holes are regular everywhere, even though
they coincide with the general relativistic ones (thus implying in particular the absence of dipolar gravitational radiation). Remarkably, this subset of the parameter space essentially
coincides with the one selected by the recent constraints from GW170817 and by solar system tests. 
\end{abstract}

\maketitle

\section{\label{sec:level1}Introduction} 

 Lorentz symmetry is believed to be a fundamental symmetry of Nature, and has been tested with high precision in a variety of settings. Indeed, violations
of Lorentz symmetry are tightly constrained in the matter sector through particle physics experiments~\cite{Kostelecky:2003fs,Kostelecky:2008ts,Mattingly:2005re,Jacobson:2005bg}, and parametrized models such 
as the Standard Model Extension~\cite{Colladay:1998fq, Kostelecky:1998id,Kostelecky:1999rh} efficiently bound such violations also in 
the interaction sector between gravity and matter~\cite{Kostelecky:2010ze}. 
Nevertheless, constraints in the gravitational sector (i.e.~from purely gravitational systems) are much less compelling. Since Lorentz 
symmetry is a cornerstone of our current understanding of fundamental physics, 
it is worth exploring ways to improve these purely gravitational constraints. One may argue that the absence of Lorentz violations (LVs) 
in the matter and matter/gravity sectors probably points to small LVs in the purely gravitational sector, but that is not necessarily the case. Indeed, mechanisms allowing
large LVs in gravity to co-exist with small LVs in matter have been put forward, and include e.g.~the emergence of Lorentz symmetry at low energies 
as a result of renormalization group running~\cite{Chadha:1982qq,Bednik:2013nxa,Barvinsky:2017kob} (see however also~\cite{Knorr:2018fdu})
 or  accidental symmetries~\cite{GrootNibbelink:2004za}, or the suppression of the percolation of LVs from gravity to matter via a large energy scale~\cite{Pospelov:2010mp}.

In order to bound LVs in gravity, one has to set up a suitable phenomenological framework. In this paper we will focus not 
on LVs tout court, but rather on violations of boost symmetry (see e.g.~\cite{Dubovsky:2004sg,Blas:2009my} for violations of spatial rotation symmetry in gravity).
A generic way to break boost symmetry is to introduce a dynamical timelike vector field (the \ae ther) defining a preferred time direction at each spacetime event.
Restricting the action to be covariant and quadratic in the first derivatives of the \ae ther, one obtains Einstein-\ae ther theory~\cite{Jacobson:2000xp}, which has been 
extensively used as a theoretical framework to understand how LVs may appear in gravitational experiments. If one further requires
that the \ae ther field not only defines a local preferred time direction, but also a preferred spacetime foliation, one ends up with 
a different Lorentz violating theory, khronometric gravity~\cite{Blas:2009qj}. The action for this theory is the same as that of Einstein-\ae ther theory 
(which is indeed the most generic action one can write at quadratic order in the derivatives), but the \ae ther field is constrained to be hypersurface orthogonal, 
i.e. parallel to the gradient of a timelike scalar field (the khronon) defining the preferred spacetime foliation.

Besides providing a theoretical framework to effectively describe LVs in gravity at low energies, khronometric theory
gains further interest from coinciding with the low energy limit of Ho\v rava gravity~\cite{Horava:2009uw}. 
The latter is a theory of gravity that is power counting~\cite{Horava:2009uw} and also perturbatively renormalizable~\cite{Barvinsky:2015kil}, 
thanks to the presence of an anisotropic scaling (Lifschitz scaling) between the time and spatial coordinates. Since this anisotropic scaling clearly breaks boost symmetry, 
Lorentz (and specifically boost) violations are crucial for the improved ultraviolet (UV) behavior of this theory.
 
Among the places where LVs play a major role is the structure of black holes (BHs). Indeed, in both Einstein-\ae ther and khronometric/Ho\v rava gravity
there exist additional graviton polarizations besides the spin-2 gravitons of General Relativity (GR). In more detail, the \ae ther vector of  Einstein-\ae ther theory
can be decomposed into spin-1 and spin-0 degrees of freedom~\cite{Jacobson:2004ts}, while the Lorentz violating khronon scalar of  khronometric/Ho\v rava gravity gives 
rise to a spin-0 polarization~\cite{Blas:2011zd}. These additional graviton polarizations propagate with speed that is generally different from 
the speed of the spin-2 modes, which in turn does not necessarily match the speed of light.\footnote{Note that the GW170817~\cite{GBM:2017lvd,PhysRevLett.119.161101}
 coincident detection of a neutron star merger in gravitational waves (GWs) and gamma rays constrains
the speed of the spin-2 mode to match almost exactly the speed of light~\cite{Monitor:2017mdv}. However, even if one includes this constraint, Lorentz violating gravity remains 
viable~\cite{Gumrukcuoglu:2017ijh}, and in particular the speed of the spin-0 mode can be very different from the speed of light~\cite{Sotiriou:2017obf}. 
We will examine in detail the experimental bounds on khronometric theory, including those from GW170817, in Sec.~\ref{sec:level2}.}
As a result, BHs have multiple horizons: a matter horizon for photons and other matter fields; 
a spin-2 horizon for tensor GWs; a spin-0 horizon for the scalar gravitational mode; and a spin-1 horizon for the gravitational vector modes, if they are present. 
Moreover, at least in spherically symmetric, static and asymptotically flat configurations, BHs also possess a universal horizon for modes of arbitrary 
speed~\cite{Barausse:2011pu,Blas:2011ni}. Modes with propagation speed diverging in the UV do indeed appear in Ho\v rava gravity when one moves away from its low energy limit (i.e. from khronometric gravity).

The regularity of these multiple event horizons has long proven a thorny issue in these theories. Already in spherical symmetry, there exists a one parameter 
family of BH solutions with regular horizons (parametrized by the mass, as in GR), but also a two parameter family of solutions (parametrized by the mass and a ``hair'' charge)
that are singular at the spin-0 horizon~\cite{Eling:2006ec,Barausse:2011pu}. Numerical simulations seem to suggest that this second family of BHs is never produced
in gravitational collapse~\cite{Garfinkle:2007bk}, but regularity becomes even more of an issue when one moves away from spherical symmetry. For instance, while slowly rotating BHs
in khronometric theory pose no particular problem~\cite{Barausse:2012qh,Barausse:2012ny,Barausse:2013nwa}, ones in Einstein-\ae ther theory generally present no universal horizon~\cite{Barausse:2015frm}. 
Moreover, they are singular at all but the outermost spin-1 horizon in regions of the parameter space of the theory's couplings where multiple spin-1 horizons 
exist~\cite{Barausse:2015frm}. There are also suggestions that the universal horizon found in static spherically symmetric BHs may be non-linearly unstable, 
at least in the eikonal (i.e. small wavelength) limit and in khronometric gravity, thus forming a finite-area curvature singularity~\cite{Blas:2011ni}. 
This may be related to the universal horizon being a Cauchy horizon in 
khronometric gravity~\cite{Bhattacharyya:2015uxt}.

To further investigate the stability and regularity of BH horizons in boost-violating gravity, we focus here on non-spinning BHs moving slowly relative to the preferred foliation in khronometric theory. This is a highly relevant physical configuration for understanding GW emission from binary systems including at least one BH. A generic feature of gravitational theories extending GR is the possible presence of dipolar gravitational radiation from quasi-circular binary systems of compact objects, e.g. neutron 
stars (c.f.~e.g.~Refs.~\cite{1975ApJ...196L..59E,0264-9381-9-9-015,PhysRevLett.70.2220,Will:1989sk,Foster:2007gr,Yagi:2013ava,Yagi:2013qpa}) 
 or BHs~\cite{Yagi:2015oca,Barausse:2016eii}. This is experimentally
very important because dipolar emission appears at -1PN order~\footnote{The post-Newtonian (PN) expansion~\cite{Blanchet:2013haa} is one in $v/c$, $v$ being the characteristic velocity of the 
system under consideration, with terms of
order $(v/c)^{2 n}$ relative to the leading one being referred to as terms of ``nPN'' order.}, i.e. it is enhanced by a factor $(v/c)^{-2}$ (with $v$ being the binary's relative velocity) compared to
the usual quadrupolar emission of GR. As such, dipolar emission may in principle 
dominate the evolution of binary systems at large separations,
a prediction that can be tested against binary pulsars data~\cite{Hulse:1974eb,Damour:1991rd}
 or the latest LIGO/Virgo detections~\cite{TheLIGOScientific:2016src,Abbott:2018lct,Barausse:2016eii}.

In Einstein-\ae ther and khronometric/Ho\v rava gravity, dipolar emission from systems of two neutron stars was studied and compared to binary pulsar observations in
Ref.~\cite{Yagi:2013qpa,Yagi:2013ava}. Refs.~\cite{Yagi:2013ava,Foster:2007gr} also laid out the theoretical framework to compute dipolar gravitational emission in these theories, showing that
the effect is proportional (as in Fierz-Jordan-Brans-Dicke theory~\cite{Fierz:1956zz,Jordan:1959eg,Brans:1961sx}) 
 to the square of the difference of the ``sensitivities'' of the two binary components~\cite{1975ApJ...196L..59E,0264-9381-9-9-015,Will:1989sk}. Ref.~\cite{Yagi:2013ava} then went on to 
 extract neutron star sensitivities from solutions of isolated stars in slow motion relative to the \ae ther/khronon. In this paper, we will follow the same program for BHs in Ho\v rava gravity, extracting their sensitivities from slowly moving solutions and drawing the implications for dipolar GW emission.

\subsection{Executive summary, layout and conventions}

The calculation of BH sensitivities turns out to be much more complicated than for neutron stars, due to the presence of multiple BH horizons and their tendency to become singular. Our main findings and conclusions can be summarized as follows:
\begin{itemize}
\item For generic values of the three dimensionless coupling constants $\alpha$, $\beta$, $\lambda$ of khronometric theory, BHs slowly moving relative to the preferred foliation present finite area curvature singularities. In more detail, if one imposes that the solution is asymptotically flat and regular at the matter horizon (which turns out to be the outermost one once experimental constraints on the theory's couplings are accounted for), a curvature singularity necessarily arises further in, at the spin-0 horizon. Giving up the requirement of asymptotic flatness allows one to obtain solutions that are regular at the spin-0 and matter horizons, but {\it not} further in, at the universal horizon, which becomes a finite-area curvature singularity.
\item If the coupling parameters of the theory are such that the speed of the spin-2 modes {\it exactly} matches that of light and the predictions of the theory in the solar system (i.e. at 1PN order) {\it exactly} match those of GR, one is still left with a one-dimensional parameter space. In more detail, these conditions set $\a=\b=0$ (which is quite natural since the experimental bounds 
on these two parameters are very tight, $|\a|\lesssim 10^{-7}$ and $|\b|\lesssim 10^{-15}$), while $\l$ can be as large as $\sim 0.01$ -- $0.1$ without violating any 
experimental constraints. In this one-dimensional subset of the parameter space,  slowly moving BHs are regular everywhere outside the central singularity at $r=0$, but coincide with the Schwarzschild solution (because the khronon profile, albeit non-trivial,
has vanishing stress energy, i.e. the khronon is a stealth field). Therefore, BH sensitivities are zero and no dipolar emission is expected from systems of two BHs. This result confirms, at the order at which we are working, the conclusion of Ref.~\cite{2014arXiv1407.1259L}, namely that khronometric theories with $\a=\b=0$ only have general
relativistic solutions in vacuum, if asymptotic flatness is imposed. We therefore expect 
GW emission to match the general relativistic predictions exactly even at higher PN orders (quadrupolar emission and higher)
if $\a=\b=0$.
\item Even if the finite area curvature singularities that we find at the spin-0 and universal horizons were due to the breakdown of our approximation scheme, and moving BHs turned out
to exist and be regular away from the central singularity at $r\neq0$, 
deviations away from the GR predictions for GW emission should be expected to be only of (fractional) order ${\cal O}[\max(\alpha,\beta)]\sim 10^{-7}$.
This is because GW generation should be exactly the same as in GR for $\alpha=\beta=0$ even at higher PN orders. Such small differences are unlikely to be observable with 
present and future GW detectors. However, if finite area curvature singularities exist (possibly smoothed by UV corrections to the low energy theory~\cite{Blas:2014aca}), 
they may give rise to ``echoes'' in the post-ringdown GW signal~\cite{Barausse:2014pra,Barausse:2014tra,Cardoso:2016rao} and/or smoking-gun features in the stochastic GW background~\cite{Barausse:2018vdb}.
\end{itemize}

The paper is organized as follows. In Sec.~\ref{sec:level2} we will briefly review Ho\v rava/khronometric gravity and the experimental constraints on its free parameters. 
In Sec.~\ref{sec:level3} we review how sensitivities of generic compact objects can be computed from slowly moving solutions, and how they are related to 
strong equivalence principle violations and more specifically to dipolar gravitational emission. In Sec.~\ref{sec:level4} we review spherical BHs 
in Ho\v rava/khronometric gravity, and introduce the ans{\"a}tze for the metric and khronon field of slowly moving BHs. In Sec.~\ref{sec:level5} we write the field equations
for slowly moving BHs and solve them for generic values of the coupling constants, while the $\a=\b=0$ case is discussed in Sec.~\ref{sec:level6}. 
Our conclusions are drawn in Sec.~\ref{sec:level7}.

Henceforth, we will set the speed of light $c=1$, and adopt a metric signature $(+,-,-,-)$.

%---------------------------------------------------------------------------------------------------------------------------------------------------------------------------------------------
%---------------------------------------------------------------------------------------------------------------------------------------------------------------------------------------------
\section{\label{sec:level2}Lorentz Violating Gravity}
%---------------------------------------------------------------------------------------------------------------------------------------------------------------------------------------------
%---------------------------------------------------------------------------------------------------------------------------------------------------------------------------------------------
In Ho\v rava gravity~\cite{Horava:2009uw}, Lorentz symmetry is violated by introducing a dynamical scalar field $T$, 
the ``khronon'', which defines a preferred time foliation. 
As such, the gradient of the khronon needs to be a timelike vector ($\nabla_\mu T\,\nabla^\mu T>0$ in our notation), 
i.e. hypersurfaces of constant khronon (the preferred foliation) are spacelike.
Using coordinates adapted to the khronon (i.e.~using $T$ as the time coordinate), 
the action for Ho\v rava gravity can be written as~\cite{Horava:2009uw,Blas:2009qj}
\be\label{action HL}
\begin{split}
S=&\frac{1-\beta}{16\pi G}\int \dd T\dd^3x\, N\sqrt{\gamma} \Big(K_{ij}\,K^{ij}-\frac{1+\lambda}{1-\beta}K^2\\
&+\frac{1}{1-\beta}{}^{(3)}{R}+\frac{\alpha}{1-\beta}a_i\,a^i+\frac{1}{M_\star^2}L_4+\frac{1}{M_\star^4}L_6\Big)\\&+S_{\rm matter}[g_{\mu\nu},\Psi]\,,
\end{split}
\ee
where $K^{ij}$, ${}^{(3)}{R}$, and $\gamma_{ij}$ are respectively the extrinsic curvature, 3-dimensional Ricci scalar and 3-metric of the 
$T=$ const hypersurfaces; $K=K^{ij}\gamma_{ij}$; $N$ is the lapse; $a_i\equiv\p_i\ln N$;  $\alpha$, $\beta$ and $\lambda$ are dimensionless 
coupling constants; and Latin (spatial) indices
are raised/lowered with the 3-metric $\gamma_{ij}$. 
The bare gravitational constant $G$ is related to the value measured
locally (e.g.~via Cavendish experiments) by~\cite{Carroll:2004ai}
\be\label{GN GHL}
G_{N}=\displaystyle\frac{G}{1-\alpha/2}\,.
\ee
The terms $L_4$ and $L_6$, suppressed by a mass scale $M_\star$, contain respectively fourth and sixth order derivatives 
with respect to the spatial coordinates, but no $T$-derivatives. Their detailed form is not needed for our purposes, 
but note that their presence is necessary to ensure power counting renormalizability of the theory.
Note that this action is \textit{not} invariant under generic 4-dimensional diffeomorphisms (exactly because it violates Lorentz symmetry) but
only under foliation-preserving diffeomorphisms
\be
T\rightarrow \tilde{T}(T)\,,\qquad x^i\rightarrow \tilde{x}^i(x,T)\,.
\ee

The matter fields, collectively denoted as $\Psi$, are assumed to couple (at the level of the action) with
the 4-metric $g_{\mu\nu}$ alone, so as to ensure that test particles move along geodesics and that
no LVs appear in the matter sector (i.e. in the Standard Model of particle physics), at least at lowest order. LVs may still 
percolate to the matter sector from the gravitational one, and suitable mechanisms suppressing this effect have therefore to be put in place 
in order to satisfy the tight bounds on LVs in the Standard Model. As already mentioned (and reviewed e.g.~in Ref.~\cite{Liberati:2013xla}), such mechanisms include for instance
the possibility that Lorentz invariance in the matter sector might 
be merely an emergent feature at low energies~\cite{Froggatt:1991ft}, due e.g.~to
renormalization group running~\cite{Chadha:1982qq,Bednik:2013nxa,Barvinsky:2017kob} or  accidental symmetries~\cite{GrootNibbelink:2004za}. 
Alternatively, as pointed out in~\cite{Pospelov:2010mp}, the matter sector and the gravitational sector
could present different levels of LVs, provided that
the interaction between them is suppressed by a high
energy-scale.

According to the precise mechanism that prevents the aforementioned percolation of LVs
from gravity to the Standard Model, the bounds on the mass scale $M_\star$ may vary. Assuming that this
percolation is efficiently suppressed, $M_\star$ needs to be $\gtrsim 10^{-2}$ eV to agree with
experimental tests of Newton's law at sub-mm scales~\cite{Blas:2010hb,Will:2014kxa}, and needs to be bound from above ($M_\star \lesssim 10^{16}$ GeV)
so that the theory is perturbative at all scales~\cite{Papazoglou:2009fj,Kimpton:2010xi,Blas:2009ck},
which is a necessary condition to apply the power-counting renormalizability arguments
of Ref.~\cite{Horava:2009uw} (see also Ref.~\cite{Barvinsky:2015kil}). 

The effect of the higher-order terms $L_2$ and $L_4$ appearing in the action \eqref{action HL} is typically 
small for astrophysical objects. Simple dimensional arguments show indeed that the fractional
error incurred as a result of neglecting those terms when studying objects of mass $M$ is
$\sim {\cal O}((G_N M M_\star)^{-2})={\cal O}(M_{\rm P}^4/ (M M_\star)^{2})$ (with $M_{\rm P}$ the
Planck mass)~\cite{Barausse:2013nwa}. Therefore, given the viable range for $M_{\star}$, the error is $\lesssim 10^{-18}(10 M_\odot/M)^2$.
For most (astrophysical) purposes, one can therefore neglect those terms, even though they are
crucial for renormalizability and for the definition of BH horizons (c.f. Refs.~\cite{Barausse:2011pu,Barausse:2013nwa}
and the discussion on universal horizons in Sec.~\ref{sec:level4}).

For these reasons, in this paper we will focus on the low-energy limit of Ho\v rava gravity, i.e.~we will neglect the 
terms $L_4$ and $L_6$  in Eq.~\eqref{action HL}. The resulting theory is often referred to
as khronometric theory. For our purposes it will also be convenient to re-write the action 
 covariantly, i.e.~in a generic coordinate system not adapted to
the khronon field, in terms of an ``\ae ther'' timelike vector $u^\mu$ of unit norm,
\be \label{khaether}
u_\mu=\frac{\nabla_\mu T}{\sqrt{\nabla^\alpha T\nabla_\alpha T}}\,.
\ee
Neglecting the $L_4$ and $L_6$ terms, the action \eqref{action HL} then becomes~\cite{Jacobson:2010mx}
\be\label{khaction}
\begin{split}
S_{\kh}=&-\frac{1}{16\pi G} \int \dd^4x\, \sqrt{-g} \Big(R+\l\;(\nabla_\m u^\m)^2\\&+\b\nabla_\m u^\n \nabla_\n u^\m+\a\; a_\m a^\m\Big)+S_{\rm matter}[g_{\mu\nu},\Psi]\,,
\end{split}
\ee
where $g$, $R$ and $\nabla$ are 4-dimensional quantities (the metric determinant, Ricci scalar and Levi-Civita connection
respectively). Note that this action is invariant under 4-dimensional diffeomorphisms, but the theory
is still Lorentz (i.e. boost) violating due to the presence of the timelike \ae ther vector $u^\mu$, which defines
a preferred time direction.

The field equations of khronometric theory are obtained by varying the action (\ref{khaction}) 
with respect to $g^{\mu\nu}$ and $T$. 
Variation with respect to the metric yields the generalized Einstein equations~\cite{Barausse:2012qh,Barausse:2013nwa}
\begin{equation}\label{khEinsteinEq}
	G_\mn- T^{\kh}_\mn=8\pi G \, T_\mn^{\rm matter}\,,
\end{equation}
where $G\lmn=R\lmn-R\, g\lmn/2$ is the Einstein tensor, 
the matter stress-energy tensor is defined as usual as 
\begin{equation}
	T^\mn_{\rm matter}=\frac{-2}{\sqrt{-g}}\frac{\delta S_{\rm matter}}{\delta g_\mn}\,,
\end{equation}
and 
the khronon stress-energy tensor is given by
\begin{equation}
\label{khSET}
\begin{split}
	&T^{{\rm kh}}_\mn\equiv \nabla_\r \left[{J_{(\m}}^\r u_{\nu)} -{J^\r}_{(\mu}u_{\nu)} -J_{(\mn)}u^\r\right]+\a\; a_\mu \, a_\nu\\
	&+\left(u_\s\, \nabla_\r J^{\r\s}-\a \,a_\rho a^\rho \right) u_\mu \, u_\nu +\frac{1}{2}L_\kh\; g_\mn+2 {\AE}_{(\mu}u_{\nu )}\,,
\end{split}	
\end{equation}
with 
\begin{align}	
&{J^\r}_\mu\equiv \l\; (\nabla_\s u^\s)\; \d_\m^\r+\b\; \nabla_\m u^\r+\a\; a_\m u^\r\,,\\
&\AE_\m \equiv  \g_\mn \(\nabla_\r J^{\r\n}-\a \, a_\r \nabla^\n u^\r \)\,,\\
& \gamma_\mn = g_\mn - u_\mu\, u_\nu\,,\\
& L_\kh= \l\;(\nabla_\m u^\m)^2+\b\nabla_\m u^\n \nabla_\n u^\m+\a\; a_\m a^\m\,.
\end{align}
Variation with respect to $T$ gives instead the scalar equation
\begin{equation}\label{khAetherEq}
	\nabla_\mu \left(\frac{\AE^\mu}{\sqrt{\nabla^\a T \nabla_\a T}}\right)=0\,.
\end{equation}
However, it can be shown that
this equation actually follows from the generalized Einstein equations~\eqref{khEinsteinEq}, from the Bianchi
identity, and from the equations of motion of matter 
(which imply in particular $\nabla_\m T^\mn_{\rm matter}=0$). This fact is also
obvious by considering diffeomorphism invariance of the covariant action~\eqref{khaction}, c.f.~Ref.~\cite{Jacobson:2010mx}.
In the following, to derive moving BH solutions, we will therefore 
solve the generalized Einstein equations \eqref{khEinsteinEq} only, in vacuum.

Moreover, in the same way in which diffeomorphism invariance implies 
the Bianchi identity in GR, diffeomorphism invariance 
of the covariant gravitational action (i.e.~Eq.~\eqref{khaction} \textit{without} the matter contribution $S_{\rm matter}$)
implies the generalized Bianchi identity:
\be\label{khBianchi}
\nabla_\mu E^\mn= \kappa \, u^\nu\,,
\ee
where we have defined
\begin{align}
&E_{\m\n}\equiv G_\mn- T^{\kh}_\mn\,,\label{khronometric Emn}\\
\label{Bianchi khronometric}
&\kappa\equiv -\frac{1}{2}\sqrt{\nabla^\a T \nabla_\a T}\; \nabla_\mu \left(\frac{\AE^\mu}{\sqrt{\nabla^\b T \nabla_\b T}}\right)\,.
\end{align}
A similar identity was derived in Ref.~\cite{Barausse:2011pu,Jacobson:2011cc} for Einstein-\ae ther theory.

%---------------------------------------------------------------------------------------------------------------------------------------------------------------------------------------------
%---------------------------------------------------------------------------------------------------------------------------------------------------------------------------------------------
\subsection{Experimental constraints}\label{ssec:Experimental constraints}

The coupling parameters $\alpha$, $\beta$ and $\lambda$ of khronometric theory need to satisfy a number of theoretical and experimental constraints, which
we will now review.

First, let us note that the theory has three propagating degrees of freedom, namely a spin-2 mode (with two polarizations) like in GR,
and a spin-0 mode. The propagation speeds of these modes in flat spacetime are respectively given by~\cite{Blas:2011zd}
\begin{subequations}\label{kh speeds}
\begin{align}
c_2^2=&\frac{1}{1-\b}\,,\\
c_0^2=&\frac{(\lambda+\beta)(2-\alpha)}{\alpha (1-\beta) (2+3\lambda+\beta)}\,.\label{spin0speed}
\end{align}
\end{subequations}
To avoid classical (gradient) instabilities and to ensure positive energies (i.e.~quantum stability, or absence of ghosts), 
one needs to impose $c_0^2>0$ and $c_2^2>0$~\cite{Blas:2011zd,Jacobson:2004ts,Garfinkle:2011iw}. 
Moreover, to prevent ultra-high energy cosmic rays from decaying into these gravitational modes
in a Cherenkov-like cascade, the propagation speeds must satisfy $c_0^2\gtrsim1-{\cal O}(10^{-15})$ and $c_2^2\gtrsim1-{\cal O}(10^{-15})$~\cite{Elliott:2005va}. GW observations also
constrain the coupling parameters and the propagation speeds. Binary pulsar observations bound the speed of the spin-2 mode to match the speed
of light to within about $0.5\%$~\cite{Yagi:2013ava,Yagi:2013qpa}, while the recent coincident detection of GW170817 and GRB 170817A~\cite{GBM:2017lvd} constrains 
$-3\times10^{-15} <c_2-1< 7\times10^{-16}$~\cite{Monitor:2017mdv}. Overall, all these constraints 
imply in particular
\be |\b|\lesssim 10^{-15}\,.
\label{doubledetection}
\ee

Further bounds follow from solar system measurements, and specifically from the upper limits on the preferred frame parameters 
$\alpha_1$ and $\alpha_2$ appearing in the parametrized PN expansion, i.e. 
$|\alpha_1|\simlt 10^{-4}$ and $|\alpha_2|\simlt 10^{-7}$~\cite{Will:2014kxa}.
Indeed, in khronometric theory these parameters are functions of the coupling constants through~\cite{Blas:2011zd,Bonetti:2015oda}
\begin{subequations}
\begin{align}
\alpha_1=&\displaystyle 4\,\frac{\alpha-2\b}{\b-1}\,,\label{solar_system1} \\
\alpha_2=&\displaystyle  \frac{\alpha_1}{8+\alpha_1}\left[1+\frac{\alpha_1(1+\beta+2\lambda)}{4(\beta+\lambda)}\right]\, .\label{solar_system2}
\end{align}
\end{subequations}
Taking into account the multi-messenger constraint (\ref{doubledetection}), for $|\lambda|\gg |\beta|$ solar system bounds thus become
\begin{subequations}
\begin{align}
4|\a| \simlt10^{-4}\,,\\
\left|\frac{\a}{\a-2}\right|\;\left|1-\a\frac{1+2\l}{\l}\right| \simlt10^{-7}\,.
\end{align}
\end{subequations}
These constraints are satisfied by $|\a|\simlt 10^{-7}$, at least if $|\lambda|\gg 10^{-7}$;
or by $|\a|\simlt 0.25\times 10^{-4}$ \textit{and} $\lambda\approx\alpha/(1-2 \alpha)$. The latter case (together with Eq.~\eqref{doubledetection}) would imply
therefore very small values for the three coupling constants, $|\alpha|\sim |\lambda|\lesssim 10^{-5}$ and $|\b|\lesssim 10^{-15}$, which seem unlikely to
allow for large observable deviations away from the general-relativistic behavior. The former case, however, while
tightly constraining $\alpha$ and $\beta$ ($|\a|\simlt 10^{-7}$, $|\b|\simlt 10^{-15}$), leaves $\lambda$ essentially unconstrained.

Indeed, the only meaningful constraint on $\lambda$ comes from cosmological observations.
For khronometric theory, the Friedmann equations take the same form as in GR, but with a gravitational constant $G_C$ different
from the locally measured one ($G_N$) and related to it by 
\be
\frac{G_N}{G_C}=\frac{2+\beta+3\lambda}{2-\alpha}\approx 1+\frac32\lambda\,,
\ee
where in the last equality we have used the aforementioned bounds on $\a$ and $\b$.
In order to correctly predict the abundance of primordial elements during Big Bang Nucleosynthesis),
which is in turn very sensitive to the expansion rate of the Universe and thus to $G_C$, one needs to impose
$|G_C/G_N-1|\lesssim 1/8$~\cite{Carroll:2004ai}. This results in $0\leq \l\lesssim 0.1$ (note that $\l$ needs
to be positive to avoid ghosts, gradient instabilities and vacuum Cherenkov radiation, as discussed at the beginning of this section; c.f. also Ref.~\cite{Yagi:2013ava}). 
Further constraints may come from other cosmological observations
(such as those of the large scale structure and the cosmic microwave background -- CMB), but have not yet been worked out in detail. 
Ref.~\cite{Audren:2013dwa} performed
some work in this direction, but required that the Lorentz violating field be the Dark Energy; the resulting bounds are therefore
inapplicable to our case. Similarly, Ref.~\cite{Afshordi:2009tt} constrained $0\leq \l\lesssim 0.01$ by using CMB observations, but assumes $\a$ and $\b$
to be exactly zero.

In summary, a viable region of the parameter space of khronometric gravity is given by $|\a|\simlt 10^{-7}$, $|\b|\simlt 10^{-15}$
and $10^{-7}\ll \l\lesssim 0.01-0.1$. This is indeed the region that we will investigate in the following.

%---------------------------------------------------------------------------------------------------------------------------------------------------------------------------------------------
\section{\label{sec:level3}Violations of the strong equivalence principle}
%---------------------------------------------------------------------------------------------------------------------------------------------------------------------------------------------
%---------------------------------------------------------------------------------------------------------------------------------------------------------------------------------------------

In theories of gravity beyond GR, the strong equivalence principle is typically violated. Indeed, such theories
generally include additional degrees of freedom besides the spin-2 gravitons of GR. Even if these additional graviton polarizations
do not couple directly to matter at the level of the action, they are typically coupled non-minimally to the spin-2 gravitons. As a result, effective
interactions between these extra gravitational degrees of freedom and matter re-appear in strong-gravity regimes, mediated by the spin-2 field 
(i.e.~by the perturbations of the metric). This effective coupling  is responsible, in particular, for the Nordtvedt effect~\cite{1975ApJ...196L..59E,PhysRev.169.1014,Nordtvedt:1968qs}, i.e. 
the deviation of the motion of binaries of strongly gravitating objects (such as neutron stars and BHs) away from the general-relativistic trajectories. 
In more detail, these deviations from GR can appear in both the conservative sector (where they can be thought of as ``fifth forces'') as well as in the dissipative one (where they can be understood as due to the radiation reaction of the extra graviton polarizations), and they strongly depend on the nature of the compact objects under consideration (e.g. whether they are neutron stars or BHs) and their properties (e.g. compactness, spin, etc).
The Nordtvedt effect has indeed been studied thoroughly in theories such as Fierz-Jordan-Brans-Dicke and other scalar tensor theories~\cite{1975ApJ...196L..59E,0264-9381-9-9-015,Will:1989sk,Barausse:2015wia,Yagi:2015oca}, and at least for neutron stars also in Einstein-\ae ther theory and khronometric gravity~\cite{Foster:2007gr,Yagi:2013ava,Yagi:2013qpa}. In this section, we will review the framework necessary to extend this treatment to the case of BHs in khronometric gravity. We refer the reader to Ref.~\cite{Yagi:2013ava} for more details.

\subsection{The sensitivities and their physical effect}

The dynamics of a compact object binary can be described in the PN approximation as long as the characteristic velocity of the system is much lower than
the speed of light~\cite{Blanchet:2013haa}.
For khronometric gravity, one has to consider two velocities, the relative velocity of the binary $v_{\rm 12}$, and the velocity of
the center of mass relative to the preferred frame $V_{\rm CM}$~\cite{Yagi:2013ava}. The former is $\ll1$ in the low-frequency inspiral phase of the binary evolution.
The latter can instead be estimated by noting that the preferred frame needs to be almost aligned with the cosmic microwave background to avoid large effects
on the cosmological evolution, hence  $V_{\rm CM}$
should be comparable to the peculiar velocity of galaxies, i.e. $V_{\rm CM}\sim 10^{-3}$.
This argument is further supported by Ref.~\cite{Carruthers:2010ii}, which showed
that the \ae ther tends to align with the time direction of the cosmological background evolution
provided that the initial misalignment (and its time derivative) are sufficiently small.

The binary components are typically described in PN theory as point particles~\cite{Blanchet:2013haa}. To account for the effective coupling to matter due to the Nordtvedt
effect, the point-particle action of GR is modified, in khronometric theory, by making the mass vary with the body's velocity relative to the preferred frame~\cite{Foster:2007gr,Yagi:2013ava}:
\be
S_{\rm pp\, A}= -\int m_A(\gamma_A)\dd\tau_A\,,
\ee
where $\dd\tau_A$ is the proper time along the body's trajectory, $\gamma_A\equiv {\bf u}_A\cdot{\bf u}$ is the projection  of the
body's four-velocity ${\bf u}_A$ on the ``\ae ther'' vector ${\bf u}$, and $A=1,2$ is an index running on the binary components.
Since both  $v_{\rm 12}$ and  $V_{\rm CM}$ are $\ll 1$, we can expand the action in $\gamma_A-1\ll 1$ as
\be\label{PPaction}
\begin{split}
S_{\rm pp\, A}= -\tilde{m}_A \int \dd\tau_A\Big\{&1+\sigma_A(1-\gamma_A)\\
&+\frac{1}{2}\sigma'_A(1-\gamma_A)^2+\mO[(1-\gamma_A)^3] \Big\}\,,
\end{split}
\ee
where $\tilde{m}_A\equiv m_A(1)$ is the body's mass while at rest with respect to the khronon, and where 
\be
\begin{split}\label{sens_def}
\sigma_A\equiv& -\frac{\dd\ln m_A(\gamma_A)}{\dd \ln\gamma_A}\Big|_{\gamma_A=1}\,,\\
\sigma'_A\equiv&\,\sigma_A+\sigma_A^2+\frac{\dd^2\ln m_A(\gamma_A)}{\dd(\ln\gamma_A)^2}\Big|_{\gamma_A=1}\,
\end{split}
\ee
are the \textit{sensitivity parameters}~\cite{Foster:2007gr,Yagi:2013ava}. These parameters encode the violations of the strong equivalence principle, and depend on the nature of the bodies
and their properties. Indeed, they can be viewed as additional ``gravitational charges'' distinct from the masses, or as ``hairs'' in the special case of BHs. 

Setting aside for the moment the problem of \textit{computing} the sensitivities, one can use the action \eqref{PPaction}, together with the modified Einstein equations~\eqref{khEinsteinEq} 
(expanded in PN orders, i.e. in $V_{\rm CM},\, v_{12}\ll 1$)
to compute the binary's motion. In particular, the sensitivities modify the conservative gravitational dynamics already at Newtonian order, i.e. the Newtonian acceleration of body $A$ is given by~\cite{Foster:2007gr,Yagi:2013ava}
\begin{equation}
\label{eq:New_active}
\dot{v}_A^i= -\frac{{\cal G} {m}_B \hat n_{AB}^i}{r_{AB}^2}\,,
\end{equation}
where  $r_{AB}=|\boldsymbol{x}_A-\boldsymbol{x}_B|$, $\hat n_{AB}^i=(x^i_A-x^i_B)/r_{AB}$, and where we have introduced the {\it{active}} gravitational masses
\be
\label{active-mass}
m_B\equiv\tilde{m}_B (1+\sigma_B)
\ee
and the ``strong field'' gravitational constant
\be
\label{calG}
{\cal G}\equiv\frac{G_N}{(1+\sigma_A) (1+\sigma_B)}\,.
\ee
The sensitivities also enter at higher PN orders in the conservative sector~\cite{Foster:2007gr,Yagi:2013ava}.

Similarly, the sensitivities also enter in the dissipative sector, i.e. in the GW fluxes. For quasi-circular orbits,
they may cause binaries of compact objects to emit dipole gravitational radiation. This effect, absent in GR (where
the leading effect is quadrupole radiation), appears at $-1$PN order, i.e. it is enhanced by a factor $(v/c)^{-2}$ relative to quadrupole radiation.
In more detail, the gravitational binding energy of a quasi-circular binary is given [because of Eq.~\eqref{eq:New_active}] by
\be
E_{b} = - \frac{{\cal{G}} \m m}{2 r_{12}}\,,
\ee
with $r_{12}$  the binary separation, $\m \equiv m_{1} m_{2}/m$ and $m \equiv m_{1} + m_{2}$,
and changes under GW emission according to the balance law~\cite{Foster:2007gr,Yagi:2013ava}
\begin{align}
\label{Pdot-AE}
\frac{\dot{E}_{b}}{E_{b}} &=  \displaystyle 2  \Big\langle  \left(\frac{ {\cal G} G \m \,m}{r_{12}^3}\right) 
  \Bigg\{ \frac{32}{5}({\cal A}_1+{\cal S}{\cal A}_2+ {\cal S}^2{\cal A}_3) v_{12}^2 \nn \\ 
  &\displaystyle+\left(s_1-s_2\right)^2\Bigg[{\cal C}+\frac{18}{5}{\cal A}_3\, V_{CM}^{j} V_{CM}^{j}\nn\\
  &\displaystyle+\left(\frac{6}{5}{\cal A}_3+36 {\cal B}\right) (V_{CM}^{i} \hat n_{12}^i)^2\Bigg]  \\ 
  &\displaystyle+\left(s_1-s_2\right)%12({\cal B}_2+2{\cal S}{\cal B}_3)V_{CM}^i \hat n_{12}^i v^j_{12} \hat n_{12}^j \nn \\&
  \displaystyle\frac{24}{5}({\cal A}_2+2{\cal S}{\cal A}_3)V_{CM}^i v_{12}^i%-2\hat n_{12}^i v_{12}^j \hat n_{12}^j
   \nn
 \Bigg\}\Big\rangle\,,
\end{align}
where we have defined the rescaled sensitivities 
\be
s_A\equiv\frac{\sigma_A}{1+\sigma_A}\,,
\ee
and we have introduced the coefficients 
\begin{align}
\label{A1-def-HL}
{\cal A}_1 &\equiv \frac{1}{c_2}+\frac{3 \a({\cal Z}-1)^2}{2 c_0(2-\a)}, \quad {\cal A}_2\equiv \frac{2({\cal Z}-1)}{(\a-2)c_0^3}, \\ 
\label{A3-def-HL}
{\cal A}_3 &\equiv \frac{2}{3\a(2-\a)c_0^5}, \quad {\cal B} \equiv \frac{1}{9\a \,c_0^5(2-\a)},\\ 
{\cal C} &\equiv \frac{4}{3 c_0^3\,\a(2-\a)},\quad {\cal S}\equiv s_1\, \frac{m_2}{m}+s_2\,\frac{m_1}{m}, \\ 
{\cal Z}&\equiv \frac{(\a_1-2\a_2)(1-\b)}{3(2\b-\a)}.
\end{align}
Note that dipole emission is proportional to the coefficient ${\cal C}$ and to the square of the difference of the sensitivities $(s_1-s_2)^2$, as in
scalar tensor theories~\cite{1975ApJ...196L..59E,0264-9381-9-9-015,Will:1989sk}.
%------------------------------------------------------------------------------------------------------------------------------------------------------------------------------------------------
\subsection{Extracting the sensitivities from the asymptotic metric}\label{ssec:Sensitivities and the asymptotic metric}

In principle, the actual values of the sensitivities for a given body (e.g. a neutron star or a BH) may be 
computed from their very definition, Eq.~\eqref{sens_def}, provided that one can obtain solutions to the field equations
for bodies in motion relative to the preferred frame, through at least order $\gamma-1={\cal O}(v^2)$, $v$ being the body's velocity in the
preferred frame (i.e. with respect to the \ae ther/khronon). Ref.~\cite{Yagi:2013ava} proposed however a simpler procedure, inspired
by a similar calculation in scalar-tensor theories~\cite{0264-9381-9-9-015}, whereby the sensitivities can be extracted from a solution
to the field equation that is accurate only through order ${\cal O}(v)$.

The idea is based on the fact that if one solves the field equations for a \textit{single} point particle [as  described by the action \eqref{PPaction}]
in motion relative to the preferred frame (or, equivalently, for a point particle at rest and a moving khronon), the sensitivity appears 
in the metric and in the khronon field near spatial infinity
already at order ${\cal O}(v)$, i.e., in a suitable gauge one has~\cite{Yagi:2013ava}
\begin{gather}\label{metric-HL-asymp_2}
\begin{split}
\ds^2 =&\displaystyle \dt^2-\dr^2+\Bigg\{-\frac{2G_N\, \tilde{m}}{r} (\dt^2+\dr^2)\\& - r^2 \left( \dth^2 +\sin^2 \th \dvph^2 \right)  \\ 
 &\displaystyle-2 v \left[ (B^{-} + B^{+}+4) \frac{G_N\, \tilde{m}}{r} \right] \cos \th \dt\, \dr \\
 & + 2v r \left[  (3+B^{-}-J) \frac{G_N\, \tilde{m}}{r} \right] \sin\th \dt\, \dth \Bigg\}\\&\times \left[1+{\cal O}\left(v,\frac1r\right)\right]\,,
 \end{split}\\
\label{HL-Foster-sph}
\begin{split}
{u}_{\m} \dx^\m =&\(\dt+v \cos\th \dr  -vr\sin\th \dth\) \\
&\times\left[1-\frac{G_N\, \tilde{m}}{r}+{\cal O}\left(\frac{1}{r^2}\right)\right]+{\cal O}(v^2)\,, 
\end{split}
\end{gather}
where $B^{\pm}$ and $J$ are defined as 
\bea
\label{B-HL-strongfield}
B^{\pm} & \equiv & \pm \frac{3}{2} \pm \frac{1}{4} (\a_1 -2\a_2) \left( 1 + \frac{2-\a}{2 \b-\a} \s \right)\nn\\
&&-\(2+\frac{1}{4} \a_1\) (1 + \sigma)\,,\\
J& \equiv&
\frac{(2+3\lambda+\b)[2(\b+ \sigma)-\a(1+ \sigma) ]}{2(\lambda+\b)(\a-2)}\,.
\eea
Therefore, the sensitivity can be read off a strong field solution valid through order ${\cal O}(v)$, i.e.~a solution describing a body
moving slowly relative to the khronon. Once such a strong-field solution is obtained, one can 
indeed extract $\sigma$ from the $g_{tr}$ and $g_{t\th}$ components of the metric, through the combinations  $3+B^{-}-J$ and $B^{-} + B^{+}+4$, respectively. 
Both readings must of course yield the same value, which we will use as a consistency check of our strong-field solution in Sec.~\ref{sec:level5}.

This was indeed the procedure used in Ref.~\cite{Yagi:2013ava} to estimate the sensitivities of neutron stars.
In the following, we will tackle the problem of finding strong-field solutions for BHs moving slowly relative to the preferred frame.

%------------------------------------------------------------------------------------------------------------------------------------------------------------------------------------------------
%------------------------------------------------------------------------------------------------------------------------------------------------------------------------------------------------

%---------------------------------------------------------------------------------------------------------------------------------------------------------------------------------------------
%---------------------------------------------------------------------------------------------------------------------------------------------------------------------------------------------
\section{\label{sec:level4}Black holes in Lorentz Violating gravity}
%---------------------------------------------------------------------------------------------------------------------------------------------------------------------------------------------
%---------------------------------------------------------------------------------------------------------------------------------------------------------------------------------------------

To construct the slowly moving BH solutions needed to extract the sensitivities, let us start from a static spherically symmetric solution
at rest relative to the khronon. We will then perturb this solution to account for the (slow) motion of the BH relative to the preferred frame.

\subsection{\label{ss:static BH}Spherical BHs at rest}

Regular (outside the central singularity at $r=0$), spherically symmetric, static and asymptotically flat BHs in khronometric theory coincide 
with those of Einstein-\ae ther theory~\cite{Jacobson:2010mx,Barausse:2013nwa} and were extensively studied in Ref.~\cite{Barausse:2011pu} (see also Ref.~\cite{Eling:2006ec}).
In Eddington-Finkelstein coordinates, their metric and \ae ther vector take the form 
\begin{gather}\label{static_metric_and_aether}
        {\rm d}\bar{s}^2= f(r)\dv^2 - 2 B(r) \dv \dr +r^2 \dOm^2\,,\\
        \bar{u}_\mu \dx\um=\frac{1+f(r)A(r)^2}{2 A(r)}\dv -A(r)B(r)\dr  \, ,\label{static_aether}
\end{gather}
where the exact functional form of the ``potentials'' $f(r)$, $B(r)$ and $A(r)$ depends on the coupling constants $\alpha$, $\beta$ and $\lambda$ and
 is obtained by solving (in general, numerically) the field equations imposing regularity at the (multiple) event horizons. 
 Note also that we have used an overbar to stress that these metric and \ae ther configurations
 will provide the background over which we will perturb in the following. 
 Because of asymptotic flatness, all three potentials asymptote to 1 at large radii, i.e.
their asymptotic solution is given by~\cite{Eling:2006ec,Barausse:2011pu} 
\begin{eqnarray}
f(r) &=& 1-\frac{2G_N\tilde{m}}{r}- \frac{\a (G_N \tilde{m})^3}{6r^3}+ \cdots \label{asyF}\\
B(r) &=& 1+\frac{\a (G_N\tilde{m})^2}{4r^2}+\frac{2\a (G_N\tilde{m})^3}{3r^3}+\cdots \label{asyB} \\
A(r) &=& 1+ \frac{G_N \tilde{m}}{r}+\frac{a_2  (G_N \tilde{m})^2}{r^2}+\nonumber\\
&&\left(24 a_2+\a-6\right) \frac{(G_N \tilde{m})^3}{12 r^3}+\cdots \label{asyA}\,,
\end{eqnarray}
where the parameter $a_2$ is determined (numerically) once the mass $\tilde{m}$ is fixed.

The causal structure of these solutions is highly non-trivial. Besides a ``matter horizon'' for photons (and in general for matter modes), defined as in GR 
by the condition $f=0$, these BHs also possess distinct horizons for the gravitational spin-0 and spin-2 modes. Since the characteristic curves of the
evolution equations for these modes correspond to null geodesics of the effective metrics~\cite{Jacobson:2004ts}
\begin{equation}\label{spinmetric}
        g^{(i)}\lab=g\lab+(c_i^2-1)u_\alpha u_\beta\, ,
\end{equation}
where $c_i$ is the propagation speed of the mode under consideration [c.f.~Eq.~\eqref{kh speeds}], the spin-0 and spin-2 horizons are defined
by the conditions $g_{\vv\vv}^{(0)}=0$ and $g_{\vv\vv}^{(2)}=0$, respectively. These horizons are typically located inside the matter horizon since the Cherenkov bound
implies $c_{0}^2,c_2^2\gtrsim1-{\cal O}(10^{-15})$.

While UV corrections -- due to the fourth and sixth order spatial derivative terms in the full Ho\v rava gravity action \eqref{action HL} --
to the metric and \ae ther solutions of Ref.~\cite{Barausse:2011pu} are negligible for astrophysical BHs (c.f. discussion
of the $L_4$ and $L_6$ terms in Sec.~\ref{sec:level2}), their presence is crucial, at least conceptually, for the causal structure of the solutions~\cite{Barausse:2011pu,Blas:2011ni}. Indeed,
because of the higher order spatial derivatives, the dispersion relations for the gravitational modes  
includes $k^4$ and $k^6$ terms ($k$ being the wavenumber), i.e. their frequency $\omega$ is given by
\be
\omega^2=c_i^2k^2+a \, k^4+b\, k^6\,,
\ee
where $a$ and $b$ are coefficients with the right dimensions. As a result, the group velocity of these modes diverges in the UV. Since matter is coupled
to the gravitational modes, similar non-linear dispersion relations will also appear in the matter sector (even though the coefficients
$a$ and $b$ are expected to be much smaller than in the gravitational sector due to the suppression of
the percolation of LVs into matter, and in general because of the weak coupling between matter and gravity). 

It would therefore appear that no event horizons should exist in the UV limit. However, Refs.~\cite{Barausse:2011pu,Blas:2011ni} identified the presence of
a ``universal horizon'' for modes of arbitrarily large speed. This horizon appears because the preferred foliation
of Ho\v rava gravity becomes a compact hypersurface in the strong field region of the BH. Modes of any speed need to move inwards at this 
hypersurface in order to move in the future preferred-time direction (defined by the preferred foliation). It can be shown~\cite{Barausse:2011pu,Blas:2011ni} that
the location of this universal horizon, which lies within the matter, spin-0 and spin-2 horizons, is defined by the condition
$u_\vv\propto 1+fA^2=0$.

Even though the exact form of the functions $f(r)$, $B(r)$ and $A(r)$ can in general be given only numerically, analytic solutions exist
in a few special cases, e.g. in the case $\alpha=0$~\cite{Berglund:2012bu}:
\begin{subequations}
\label{fAB_alpha0}
\begin{align}
        f(r) &= 1 - \frac{2 G_N \tilde{m}}{r} - \frac{\beta r_{\rm kh}^4}{r^4}, \,\,\,\,\, B(r) = 1\,,\label{fAB_alpha0_a}\\
        A(r) &= \frac{1}{f}\left(-\frac{r_{\rm kh}^2}{r^2}+ \sqrt{f+\frac{r_{\rm kh}^4}{r^4}}\right),\label{fAB_alpha0_b} \\
	r_{\rm kh} &= \frac{G_N \tilde{m}}{2}\left(\frac{27}{1-\beta}\right)^{1/4}\label{fAB_alpha0_c}
\end{align}
\end{subequations}
It can be easily checked that the universal horizon
and the spin-0 horizon coincide in this particular case [since when $\alpha\to 0$ the spin-0 speed given by Eq.~\eqref{spin0speed} diverges],
and are both located at $\ruh = {3}G_N \tilde{m}/2$. Note also that this solution does not depend on the coupling parameter $\lambda$, even though
that is \textit{not} assumed to vanish.

In the following, we will use spherically symmetric, static and asymptotically flat BHs as the starting point for the construction of our slowly moving solutions.
These spherical BHs are either produced numerically as in Ref.~\cite{Barausse:2011pu}, or given by the explicit solution \eqref{fAB_alpha0} for $\alpha=0$.

%------------------------------------------------------------------------------------------------------------------------------------------------------------------------------------------------
%------------------------------------------------------------------------------------------------------------------------------------------------------------------------------------------------
\subsection{Slowly moving BHs}\label{ss:Construction of the Ansatz}

Let us now construct ans{\"a}tze for the metric and khronon field of a (non-spinning) BH moving slowly relative to the preferred frame, based on the symmetries
of the problem. Let us first place ourselves in the reference frame comoving with the BH, i.e.~consider the physically equivalent situation where the BH is actually at rest, 
while the khronon (which determines the preferred frame) is moving relative to it with
small velocity $-v^i$ along the $z$-axis~\footnote{Note the different script that differentiates this velocity from the coordinate time $\rm v$.}.
In order for the metric to be asymptotically flat, one will therefore have to impose $g_{\mu\nu}=\eta_{\mu\nu}+{\cal O}(1/r)$ 
and $u^\mu\partial_\mu= \partial_t-v \partial_z+{\cal O}(v)^2$ in the Cartesian coordinates $(t,x^i)$.

To exploit the symmetry of the configuration under rotations around the $z$ axis, it is convenient to adopt cylindrical isotropic coordinates $(t,\r,z,\phi)$, in which
the background ${\cal O}(v)^0$ spherical BHs of Sec.~\ref{ss:static BH} can be written as
\begin{gather}
{\rm d}\bar{s}^2=f(r(\tilde{r}))\dt^2-b^2(\tilde{r})\left(\dd\rho^2+\r^2\dd\phi^2+\dz^2\right)\,,\\
\bar{u}_\m \dd x^\m=A(r(\tilde{r})) \dt + \bar{u}_{\tilde{r}}(\tilde{r}) \dd\tilde{r}\,.
\end{gather}
Here, $\bar{u}_{\tilde{r}}$ is determined by the normalization condition $u_\m u^\m=1$; $\tilde{r}=\sqrt{\r^2+z^2}$ is the radial isotropic coordinate, which is related to
the areal radius $r$ used in Eqs.~\eqref{static_metric_and_aether} and \eqref{static_aether} by the relation $r=\tilde{r}\,b(\tilde{r})$; and $b(\tilde{r})$ is related to $B(r)$ by the relation
\be
\frac{B(r)}{\sqrt{f(r)}}= \frac{b(\tilde{r})}{b(\tilde{r})+\tilde{r}\;  \dd b(\tilde{r})/\dd\tilde{r}}\,.
\ee
Also note that the time coordinate $t$ is related to the Eddington-Finkelstein time coordinate $\rm v$ by $t={\rm v}-\int^{r(\tilde{r})}_{\bar{r}} B(r)/f(r) \dr$,
where $\bar{r}$ is a reference radius.

The use of isotropic coordinates makes it simple to construct the ans{\"a}tze for the ${\cal O}(v)$ perturbations. Following the idea briefly outlined
in Appendix A of Ref.~\cite{Yagi:2013ava} for stellar systems, we can observe that the perturbations $\delta g_{tt}$ and $\delta u_t$ transform as scalars under spatial rotations;
$\delta g_{ti}$ and $u_i$ transform as vectors; and $\delta g_{ij}$ transforms as a tensor. Since we only have two 3-vectors, $v^i$ and $n^i=x^i/|x|$, to construct these
quantities, we can write, without loss of generality,
\begin{subequations}\label{geometric ansatz}
\begin{align}
\dg_{tt}=&\, \a_1(\tir)\;\vec{n}\cdot\vec{v}\,,\\
\d u^t= &\,\b_1(\tir)\; \vec{n}\cdot\vec{v}\,,\\
\left(
\begin{array}{c}
\dg_{t\rho} \\ \dg_{tz} \end{array}\right)
=&\, \a_2(\tir) (\vec{n}\cdot\vec{v})\vec{n}+\a_3(\tir)\vec{v},\\
\left(\begin{array}{c} \d u^\rho \\ \d u^z \end{array} \right)=&\,\b_2(\tir)\,(\vec{n}\cdot\vec{v})\vec{n}+\b_3(\tir)\vec{v}, \\
\left(\begin{array}{cc}
\dg_{\rho\rho} & \dg_{\rho z}\\
\dg_{z\rho} & \dg_{zz}\end{array}\right)=& \,\a_4(\tir) (\vec{n}\cdot\vec{v}) \vec{n}\otimes\vec{n}\nn\\&\,+\a_5(\tir)(\vec{n}\otimes\vec{v}+\vec{v}\otimes\vec{n})\,,\\
\d u^\phi=\dg_{r\phi}=\,&\dg_{\phi\phi}=\dg_{\r\phi}=\dg_{z\phi}=0\,,
\end{align}
\end{subequations}
where we have introduced the potentials $\a_{i}(\tir)$ for $i=1,2,3,4,5$ and $\b_i$ with $i=1,2,3$, which must depend only on the radial
coordinate $\tilde{r}$ (and not on $\r$ and $z$ singularly) to ensure the right transformation properties under rotations. Note
that actually only six of these eight potentials are independent, as the (perturbed) \ae ther $u^\mu=\bar{u}^\mu+\delta u^\mu$
must satisfy the normalization condition ${u}^\mu {u}_\mu=1$ and be hypersurface orthogonal, i.e. $\epsilon^{\mu\nu\a\b}{u}_\nu \partial_\a {u}_\b=0$ [c.f.~Eq.~\eqref{khaether}].
Also note that $\delta u^\phi$, $\dg_{t\phi}$, $\dg_{\phi\phi}$, $\dg_{\r\phi}$ and $\dg_{z\phi}$ must vanish because neither $v^i$ nor $n^i$ 
possesses a tangential component in the $\phi$
direction. (One may in principle obtain non-zero values for these components by introducing the tangential pseudovector $\vec{n}\times \vec{v}$, but that
would violate parity, which would be incompatible with the symmetries of the system, which does not rotate around the $z$-axis.)

Transforming now back to the original Eddington-Finkelstein coordinates that we will use in this paper, the most generic
form of the metric and \ae ther vector then becomes
\begin{align}
\label{metric_Ansatz}
{g}_{\mu\nu} dx^\mu dx^\nu =
& f(r) \dv^2 -2B(r) \dr \dv-r^2 \dOm^2\nn\\
&+v\,\Big\{{\dv}^2 f(r)^2 \cos \theta \psi (r)\nn\\
&-2 {\dd\theta } {\dr} \sin \theta [\Sigma (r)-B(r) \chi (r)]\nn\\
&+2 {\dr}  {\dv} f(r) \cos \theta [\delta (r)-B(r) \psi (r)]\nn\\
&+{\dr}^2 B(r) \cos \theta [B(r) \psi (r)-2 \delta (r)+2 \Delta (r)]\nn\\
&-2 {\dd\theta } {\dv} f(r) \sin \theta \chi (r)\Big\} +\mO(v^2)\,, 
\end{align}
and
\begin{align}\label{aether_Ansatz}
{u}_\m \dx^\m=&\bar{u}_\vv(r)\dv -A(r)B(r)\dr+v\,\Bigg\{ \frac{1}{2} f(r) \cos \theta \times\nn\\
& \Bigg[2 \bar{u}^r(r) \left(\frac{B(r) \Delta (r) \bar{u}^r(r)}{\bar{u}_\vv(r)}+\delta (r)-\eta (r)\right) \nn\\
   &+\psi (r) \bar{u}_\vv(r)\Bigg] \dv+\frac{1}{2} \cos \theta \times \nn\\
   &\Bigg[B(r) \Big(-\frac{2 B(r) \Delta (r) \bar{u}^r(r)^2}{\bar{u}_\vv(r)}-2 \delta (r) \bar{u}^r(r)\nn\\
   &-\psi (r) \bar{u}_\vv(r)\Big)+2 A(r)f(r) \eta (r)\Bigg] \dr\nn\\
   &-\sin \theta \Pi (r) \bar{u}_\vv(r) d\theta\Bigg\}+\mO(v^2)\,,
\end{align}
where the background \ae ther components $\bar{u}_\vv$ and $\bar{u}^r$ are given by Eq.~\eqref{static_aether}, i.e. $\bar{u}_\vv=(1+fA^2)/(2A)$ and $\bar{u}^r=(-1 + A^2 f)/(2 AB)$,
\be
\begin{split}
\eta(r)=&-\frac{2\bar{u}^r(r)^3 B(r)^3 \Delta (r) -2 \bar{u}_\vv(r)^3 \Pi '(r)}{2 f(r) \bar{u}_\vv(r)}\\
&-\frac{B(r)^2 \bar{u}_\vv(r) \bar{u}^r(r) \left[2 \delta(r) \bar{u}^r(r)+\psi (r) \bar{u}_\vv(r)\right]}{2 f(r) \bar{u}_\vv(r)}
\end{split}
\ee
to ensure hypersurface orthogonality, and the six independent potentials $\delta,\chi,\psi,\Delta,\Sigma,\Pi$ 
are algebraically related to the potentials $\a_{i}$ and $\b_i$  introduced above.
This ansatz can then be further simplified by noting that a gauge transformation
with generator $\xi^\mu \partial_\mu=\Omega(r)(-r  \cos\theta\partial_r+ \sin\theta\partial_\theta)$
can be used, by choosing the function $\Omega(r)$ appropriately, to set any one of six potentials (e.g. $\Delta$) to zero,
while leaving the form of the ansatz \eqref{aether_Ansatz} unchanged (modulo redefinitions of the remaining potentials). 
By performing then a gauge transformation $\vv'=\vv-v\,\Pi(r)\cos\th+\mO(v^2)$
one can also send $\Pi$ to zero. 
In the following, we will therefore set $\Delta=\Pi=0$.

One is therefore left with four independent potentials $\delta,\chi,\psi,\Sigma$,
which near spatial infinity ($r\to+\infty$) must satisfy the boundary conditions
$\psi,\Sigma \to 0$, $\delta\to-1$ and $\chi/r\to -1$ in order to ensure asymptotic
flatness.
Indeed, it is easy to see that these conditions lead to $\ds^2\approx \dt^2-\dr^2-r^2\dd\Omega^2+2v \dt \dz$,
where we have changed time coordinate to $t\approx\vv-r$ and $z=r\cos\theta$.
A further coordinate change $t'=t+vz$ 
transforms the line element into the flat one. As for the \ae ther, the same coordinate transformations yield
${u}^\m \partial_\mu\approx \partial_t-v\partial_z$ asymptotically, i.e. near spatial infinity the \ae ther moves with velocity $-v$
relative to the flat asymptotic metric. Another way of checking these boundary conditions is to note that they correspond to $\beta_3\to-1$
and $\beta_{1,2},\alpha_{1,2,3,4,5}\to 0$ in terms of the potentials introduced in Eqs.~\eqref{geometric ansatz}.

As expected from the symmetries of the problem (see also Appendix A of Ref.~\cite{Yagi:2013ava}) the field equations \eqref{khEinsteinEq}, when evaluated with these
ans{\"a}tze, become ordinary differential equations in the radial coordinate, i.e. the dependence on the polar angle $\theta$ drops out.
This is a highly non-trivial (but expected, if our ans{\"a}tze is correct) fact  that simplifies the search for solutions,  to be compared for instance
with the procedure followed by Ref.~\cite{Yagi:2013ava}, which involved projecting the field equations onto Legendre polynomials. It is 
also useful as an a posteriori check of our calculations.

%------------------------------------------------------------------------------------------------------------------------------------------------------------------------------------------------
\section{Field equations and numerical solutions \label{sec:level5}}

In this section, we will first analyze the structure of the vacuum field equations 
for the ${\cal O}(v)$ potentials $\d(r)$, $\c(r)$, $\psi(r)$ and $\S(r)$ introduced in the previous section. 
We  will  then analyze the boundary and regularity conditions that those potentials must satisfy, and 
obtain numerical solutions for them under various choices of those conditions.

\subsection{Structure of the field equations}

By replacing the metric and \ae ther ans{\"a}tze, (\ref{metric_Ansatz}) and (\ref{aether_Ansatz}), into the vacuum field equations $E^{\mu\nu}=\bar{E}^{\mu\nu}+\delta E^{\mu\nu}=0$ 
and expanding in $v$,
one obtains ordinary differential equations for the background potentials $f$, $A$ and $B$ at zeroth order, and for $\d(r)$, $\c(r)$, $\psi(r)$ and $\S(r)$ at first order. 
Since the background solutions are known from previous work (c.f.~Sec.~\ref{ss:static BH}), we will focus here on the first order equations.

Naively, there appear to be six non-trivial field equations at first order, coming from the perturbations $\delta{E^\vv}_\vv$, $\delta{E^\vv}_\th$, $\delta{E^r}_\vv$, $\delta{E^r}_r$, $\delta{E^r}_\th$, and $\delta{E^\th}_\th$ of Eq.~\eqref{khronometric Emn}. However, because of the generalized Bianchi identity Eq.~\eqref{khBianchi}, only four of these equations are actually independent, thus providing a closed problem for the potentials $\d(r)$, $\c(r)$, $\psi(r)$ and $\S(r)$. In 
more detail, Eq.~\eqref{khBianchi} has three non-trivial components through linear order in $v$ (the $\phi$ component being trivial since
both sides of the identity are ${\cal O}(v)^2$). Since we are not solving the khronon equation \eqref{khAetherEq} [because that is automatically 
implied by the modified Einstein equations, as discussed in Sec. \ref{sec:level2} and as can also be seen, at least in vacuum, from the identity Eq.~\eqref{khBianchi} itself], it is convenient
to eliminate $\kappa$ from the three non-trivial components of Eq.~\eqref{khBianchi}. This leads to the identities
\begin{subequations}
\begin{align}
u_r\nabla_\m {E\um}_\vv-u_\vv\nabla_\m {E\um}_r={\cal O}(v^2)\,,\\
u_\th\nabla_\m {E\um}_r-u_r\nabla_\m {E\um}_\th={\cal O}(v^2)\,,
\end{align}
\end{subequations}
which can in turn be rewritten as
\begin{align}
&\nabla_\m(u_r \, {E\um}_\vv-u_\vv \, {E\um}_r)=\nn\\&\qquad {E\um}_\vv \nabla_\m u_r  -{E\um}_r \nabla_\m u_\vv+{\cal O}(v^2)    \,,\\
&\nabla_\m {E\um}_\th={\cal O}(v^2)\,.
\end{align}
where we have used (in the second equation) the fact that $u_\th={\cal O}(v^2)$ in our gauge.

By expanding the summations in these identities, it is clear that the ${E^r}_\th$ and the combination $u_r \, {E^r}_\vv-u_\vv \, {E^r}_r$ 
must depend on the potentials $f$, $A$ and $B$ (at zeroth order) and $\d$, $\c$, $\psi$ and $\S$ (at first order) 
through \textit{one less} radial derivative than the highest derivatives appearing in the rest of the field equations.
Moreover, from the same (expanded) identities it follows that these two quantities are initial value constraints
for evolutions in the radial coordinate, i.e. if they are set to zero at some finite radius, they remain zero at all
other radii \textit{if} the remaining field equations (the ``evolution equations'') are solved. As can be seen, this follows from the generalized Bianchi
identity in the same way in which in GR the Bianchi identity allows for splitting initial value problems in energy and momentum constraints 
and evolution equations. The same procedure was also followed in Ref.~\cite{Barausse:2011pu} to split the field equations of Einstein-\ae ther gravity
into constraints and evolution equations (in the radial coordinate) in static spherically symmetric configurations.

The explicit equations for $\d$, $\c$, $\psi$ and $\S$  are too complicated to be presented here,
but their schematic form is given as follows. The evolution equations have the following structure:
\begin{subequations}\label{potential dynamical equations}
\begin{align}
e_1\equiv&\d''(r)-\sum\limits_{n=1}^7 w^\delta_n(r) M_n   \label{potential dynamical equations delta}=0\,,\\
e_2\equiv&\c''(r)-\sum\limits_{n=1}^7 w^\c_n(r) M_n  \label{potential dynamical equations chi}=0\,,\\
e_3\equiv&\psi''(r)-\sum\limits_{n=1}^7 w^\psi_n(r) M_n \label{potential dynamical equations psi}=0\\
e_4\equiv&\S'(r)-\sum\limits_{n=1}^7 w^\S_n(r) M_n \label{potential dynamical equations Sigma}=0\,,
\end{align}
\end{subequations}
where $\vec{M}\equiv[\d(r),\c(r),\psi(r),\S(r),\d'(r),\c'(r),\psi'(r)]$ and $w^\delta_n(r),w^\c_n(r),w^\psi_n(r),w^\S_n(r)$ (with $n=1,\ldots,7$) are functions
of the radial coordinate, the background solution $f,A,B$ and the coupling constants. As for the constraints $C_1(\vec{M})$ and $C_2(\vec{M})$, they satisfy
the conservation equations
\begin{equation}\label{constraints}
\frac{\dd C_i}{\dr}=\sum\limits_{n=1}^2 w^{C_i}_n(r) C_n+\sum\limits_{n=1}^4 w^{e,C_i}_n(r)\, e_n\,,\quad i=1,2\,,\\
\end{equation}
where again the coefficients $w^{C_i}_n(r)$ and $w^{e,C_i}_n(r)$ (with $i=1,2$ and $n=1,\ldots,4$)
depend also on the background solution and the coupling constants. Note that at least at large radii, the coefficients $w^{C_i}_n(r)$ are negative,
which contributes to damping potential violations of the constraints during our radial evolutions. 

Finally, let us also note that because Eqs.~\eqref{potential dynamical equations}--\eqref{constraints} are linear and homogeneous, 
one is free to rescale any one solution by a constant factor, i.e. given a solution
$[\d(r),\c(r),\psi(r),\S(r)]$, also $\Lambda[\d(r),\c(r),\psi(r),\S(r)]$, with $\Lambda=$ const, is a solution. 
We will use this fact when setting the initial/boundary conditions for the system given by Eqs.~\eqref{potential dynamical equations} in the next section.

\subsection{Solutions regular at the matter horizon}\label{ss:Boundary conditions}

Before solving the system given by Eqs.~\eqref{potential dynamical equations}, let us comment on the boundary/initial  conditions that the solution needs to satisfy.
Close inspection of the coefficients $w^\delta_n(r),w^\c_n(r),w^\psi_n(r),w^\S_n(r)$ 
shows that the system presents at least three potentially singular points (with $r\neq0$) where at least one of the coefficients diverges. These three singularities are located
at the matter horizon, at the spin-0 horizon, and at the universal horizon. Regularity at these radial positions needs therefore to be enforced. On top of this, 
physically relevant solutions should asymptote to flat space and to a khronon moving with velocity $-v$ near spatial infinity, which translates into
the boundary conditions $\psi,\Sigma \to 0$, $\delta\to-1$ and $\chi/r\to -1$ as $r\to\infty$ as shown in Sec.~\ref{ss:Construction of the Ansatz}.

Let us first attempt to impose regularity at the outermost of these positions, the matter horizon. If the potentials are regular there\footnote{One can show that analyticity of
the potentials $\delta$, $\chi$, $\psi$ and $\Sigma$ is required to ensure finiteness of the invariants constructed with
the metric, the \ae ther vector, and the Killing vectors $\partial_\vv$
and $\partial_\phi$ (e.g. $R$, $R_{\mu\nu} R^{\mu\nu}$, $R_{\mu\nu\alpha\beta} R^{\mu\nu\alpha\beta}$, and scalars obtained
by contracting among themselves curvature tensors, Killing vectors and the \ae ther).}, they can be Taylor-expanded
as
\begin{subequations}\label{mh_initialconditions}
\begin{align}
\d(r)=&\sum\limits_{k=0}^\infty \d_{k,\rm h}\,(r-r_{\rm h})^k\,,\\
\c(r)=&\sum\limits_{k=0}^\infty \c_{k,\rm h}\,(r-r_{\rm h})^k\,,\\
\psi(r)=&\sum\limits_{k=0}^\infty \psi_{k,\rm h}\,(r-r_{\rm h})^k\,,\\
\S(r)=&\sum\limits_{k=0}^\infty \S_{k,\rm h}\,(r-r_{\rm h})^k\,,
\end{align}
\end{subequations}
where $r_{\rm h}$ is the matter horizon's position, and the coefficients $\d_{k,\rm h}$, $\c_{k,\rm h}$, $\psi_{k,\rm h}$ and $\S_{k,\rm h}$ must be 
determined by solving the field equations. Indeed, solving the evolution \textit{and} constraint equations perturbatively near $r=r_{\rm h}$ allows one to express all those coefficients
as a function of $\delta_{0,\rm h}$ and $\Sigma_{0,\rm h}$ alone, i.e. the solution only has two independent degrees of freedom near the matter horizon. Those
can then be reduced to just one by rescaling the solution by a constant factor as described in the previous section, whereby one can set e.g. $\Sigma_{0,\rm h}=1$
and keep $\delta_{0,\rm h}$ free.\footnote{Rescaling $\Sigma_{0,\rm h}=1$ is only possible if $\Sigma_{0,\rm h}\neq0$. However, setting $\Sigma_{0,\rm h}=0$
does not allow for a solution that is asymptotically flat when integrating outwards.}

This parameter then needs to be determined by imposing asymptotic flatness. We therefore use the perturbative solution \eqref{mh_initialconditions} to move slightly away from $r=r_{\rm h}$, and then integrate numerically the system given by Eqs.~\eqref{potential dynamical equations} up to large radii. The value of $\delta_{0,\rm h}$
is then determined by imposing that $\delta$ and $\chi/r$ asymptote to the (same) constant (which does not need to be -1, because
we have rescaled the solution by a global unknown factor) and that $\psi,\Sigma \to 0$ at large radii.
In practice, we perform a bisection procedure on the value of $\delta_{0,\rm h}$, according to whether $\delta(r)$ diverges to positive or negative values
as $r\to\infty$. This is similar to what was done in Ref.~\cite{Barausse:2011pu} for the static, spherically symmetric and asymptotically flat solutions that we employ as our background.
Ref.~\cite{Barausse:2015frm} also used a similar procedure to find slowly rotating BHs in Einstein-\ae ther theory.

In more detail, solving the evolution equations perturbatively near spatial infinity 
and assuming that $\delta(r)$ asymptotes to a constant there, one finds the asymptotic solution
\begin{subequations}\label{asymptotic}
\begin{align}
\d(r)=&\,\d_{0}-\frac{2(\b+\l)(G_N\tilde{m}\d_0+2\c_0)}{(1-3\b-2\l)r}+\mO\left(\frac{1}{r^2}\right)\,,\label{asymptotic delta}\\
\c(r)=&\,\d_0\, r+\c_0+\mO\left(\frac{1}{r}\right)\,, \label{asymptotic chi}\\
\psi(r)=&\,\frac{3\b (3-2a_2)(G_N\tilde{m})^2\d_0-2\S_1}{3\, r^2}+\mO\left(\frac{1}{r^3}\right)\,,\\\
\S(r)=&\,\frac{\S_1}{r}+\mO\left(\frac{1}{r^2}\right) \,,
\end{align}
\end{subequations}
where $\d_0$, $\c_0$, $\c_2$ and $\Sigma_1$ are free parameters that can be determined from our numerical solutions, once the bisection has converged.
The constraint equations are also satisfied by this asymptotic solution.
Note that we use the relations between the coefficients of Eq.~\eqref{asymptotic} to \textit{test} a posteriori the consistency of our numerical solutions.

Note also that if one inserts the solution \eqref{asymptotic} into the metric ansatz (\ref{metric_Ansatz}), the resulting metric can
be put in the same gauge as Eq.~(\ref{metric-HL-asymp_2}) by first transforming Eddington-Finkelstein to Schwarzschild coordinates,
and by then applying an infinitesimal coordinate transformation with generator $\delta t\propto v r \cos\th+{\cal O}(v)^2$, 
$\delta r\propto  -v\, \Omega(r) \cos\th  +{\cal O}(v)^2$ and $\delta \theta\propto  v\, \Omega(r) \sin\th/r  +{\cal O}(v)^2$, with $\Omega(r)={\cal O}(1/r)$ a
suitable function. By comparing the metric obtained in this way to Eq.~(\ref{metric-HL-asymp_2}), one can then relate the sensitivity $\sigma$
to the two parameters $\d_0$ and $\c_0$:
\be\label{sensitivity}
\begin{split}
\sigma= &\frac{\a-\b-3\a \b+5\b^2+\l-2\a \l +3\b \l }{(2-\a)(1-3\b-2\l)} \\&+\frac{2(1-\b)(\b+\l)}{(2-\a)(1-3\b-2\l)}\frac{\c_0}{G_N \tilde{m}\d_0}\,.
\end{split}
\ee
We have also checked this equation by solving directly the field equations near spatial infinity for a point particle described by the action \eqref{PPaction}, 
in the gauge of the metric ansatz Eq.~(\ref{metric_Ansatz}), and then by comparing to the asymptotic solution given by Eq.~\eqref{asymptotic}.
Note that this is the same procedure (although in a different gauge) as that followed by Ref.~\cite{Yagi:2013ava} to relate sensitivities
to coefficients appearing in the asymptotic metric and \ae ther vector of isolated neutron stars.

Our numerical solutions confirm that if one imposes regularity at the matter horizon and asymptotic boundary conditions corresponding to a flat spacetime and a khronon 
moving with velocity $-v$ relative to the BH, the sensitivities are non-vanishing. We have not performed a systematic investigation of the viable region 
of the parameter space described in Sec.~\ref{ssec:Experimental constraints}, because of the difficulty of obtaining numerical background solutions
for the potentials $f$, $A$ and $B$ for small but non-zero value of $\alpha$ and $\beta$. We will nonetheless study later, in Sec.~\ref{sec:level6}, solutions
for $\alpha=\beta=0$ and $\lambda\neq0$, and extract their sensitivities. For the moment, let us mention that for values of $\alpha\sim\beta\sim 10^{-2}$ and $\lambda\sim 0.1$, we obtain $\sigma\sim 10^{-3}$.

\begin{figure}[thbp]
\centering
\includegraphics[width=0.495\textwidth]{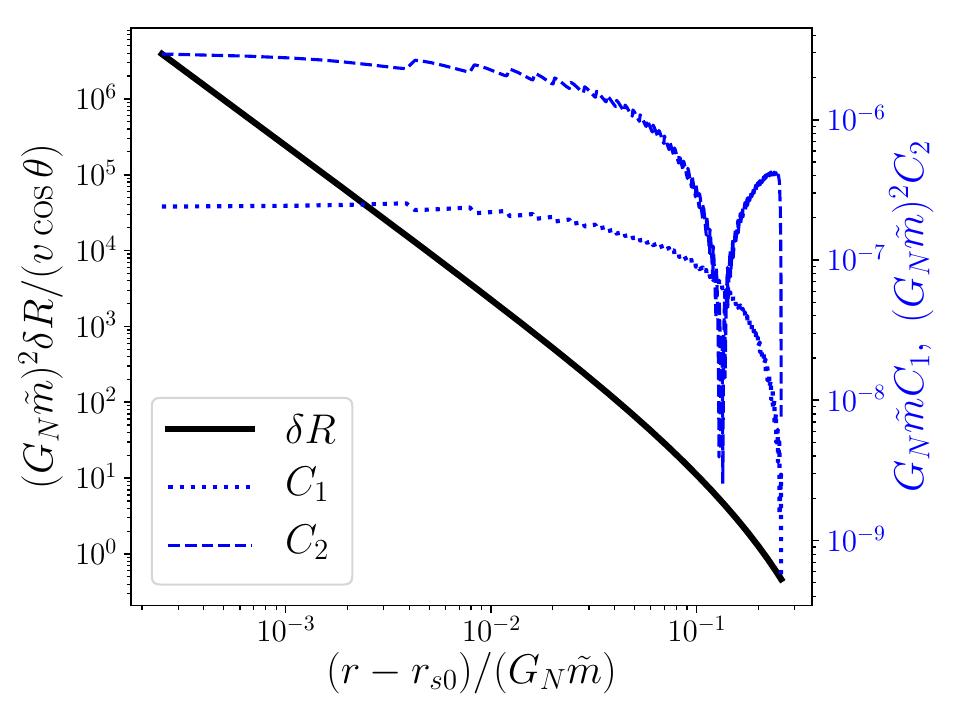}
\caption{${\cal O}(v)$ contribution to the Ricci scalar near the spin-0 horizon (left axis) and constraint violations (right axis) as a function
of distance from the spin-0 horizon, 
for the asymptotically flat solution regular at the matter horizon, and for $\a=0.02$, $\b=0.01$ and $\l=0.1$.}
\label{spin-0 divergence}
\end{figure}

One important caveat, however, is that it is not at all clear that these solutions (and the corresponding values of the sensitivities) are 
physically significant, as the numerical solutions that we obtain diverge when integrating inwards from the matter horizon to the spin-0 horizon.
We have also checked that this divergence extends to the curvature invariants, i.e. these solutions seem to present
a finite-area curvature singularity at the spin-0 horizon.
Indeed, when we integrate inwards the asymptotically flat and regular (at the matter horizon) solution, 
we find that the curvature invariants diverge at the spin-0 horizon, already in regions
where our numerical scheme is not yet breaking down. This is shown in Fig.~\ref{spin-0 divergence}, 
which plots the fastest growing curvature invariant
of the geometry,  
as well as the constraint violations occurring in the numerical integration.

Because all free parameters of the solution were determined by imposing regularity at the metric horizon and by the boundary conditions at 
spatial infinity, the spin-0 horizon curvature singularity, which was already visible in the field equations \eqref{potential dynamical equations} and \eqref{constraints}, seems
almost unavoidable, and reminiscent of similar finite-area curvature singularities appearing at all but the outermost spin-1 horizons of slowly rotating BHs in Einstein-\ae ther theory,
for coupling values allowing for such multiple spin-1 horizons~\cite{Barausse:2015frm}.
We will further investigate this singularity, and in particular whether it can be avoided thanks to a field redefinition, in the next section.

\subsection{Solutions regular at the matter and spin-0 horizon}

Curvature singularities at the spin-0 horizon also appear when studying spherical BHs in Ho\v rava gravity and Einstein-\ae ther theory. 
Refs.~\cite{Eling:2006ec,Barausse:2011pu} found a two-parameter family of asymptotically flat, static and spherically symmetric BH solutions in those theories. One of the two free parameters is
the mass of the BH, while the second is a ``hair'' regulating whether the spin-0 horizon is singular or not.
Indeed, after imposing regularity at the matter horizon, for generic values
of this parameter the spin-0 horizon is singular, and regularity at that location is obtained only for one specific, ``tuned'' value of that parameter. (That value is a function of
the mass and the coupling constants of the theory.) 
As argued in the previous section, in our case we have no free parameter to tune to impose regularity at the spin-0 horizon, which we therefore expect to be truly singular.

 To verify even further the existence of a curvature singularity at the spin-0 horizon, one can follow Refs.~\cite{Eling:2006ec,Barausse:2011pu} and note 
 that the action~\eqref{khaction} is invariant under the field redefinition~\cite{Foster:2005ec}
\be\label{eq:field redefinition}
g_{\m\n} '= g_{\m\n}+(\zeta-1)u_\m u_\n\,,\quad T'=T\,,
\ee
where $\zeta$ is a constant, provided that the original  $\a$, $\b$ and $\l$ are replaced by $\a'$, $\b'$ and $\l'$ 
satisfying
\be\label{ci_scaling}
\begin{split}
\a'=&\,\a\,,\\
\b'+\l'=&\,\zeta \, (\b+\l)\,,\\
\b'-1=&\,\zeta (\b-1)\,.\\
\end{split}
\ee
Choosing in particular $\zeta=c_0^2$, the redefined metric $g'$ coincides with the spin-0 metric [c.f. Eq.~\eqref{spinmetric}]. This
therefore allows one to cast the original problem, characterized by the metric $g$ and the couplings $\a$, $\b$, $\l$, into
one involving the spin-0 metric $g'=g^{(0)}$ and the new couplings $\a'$, $\b'$ and $\l'$. The advantage of this ``spin-0 frame'' is
that the matter and spin-0 horizons now coincide (as they are both defined in terms of characteristics of the metric $g'=g^{(0)}$, 
i.e. by the condition $g'_{\rm vv}=0$ in spherical symmetry), so one can easily impose regularity at both. This is indeed the way 
Refs.~\cite{Eling:2006ec,Barausse:2011pu} impose regularity at both the matter and spin-0 horizon in the spherical static case.

Working therefore in the spin-0 frame, we impose regularity at the matter/spin-0 horizon location $r_h$ by solving the evolution and constraint equations
perturbatively with the ansatz given by Eq.~\eqref{mh_initialconditions}. (Analyticity of
the potentials $\delta$, $\chi$, $\psi$ and $\Sigma$ is again required to ensure finiteness of the invariants constructed with
the metric, the \ae ther vector, and the Killing vectors.) The number of free parameters of the resulting solution is however
different than what was obtained in Sec.~\ref{ss:Boundary conditions}. This is because in the spin-0 frame one has $c_0=1$ (this can be verified explicitly by using the new
coupling parameters given by Eq.~\eqref{ci_scaling}, with $\zeta=c_0^2$, into Eq.~\eqref{spin0speed}), which changes the structure of the equations,
because of the presence of factors $c_0^2-1$ in the denominators. (The explicit form of the equations is again too long and cumbersome to show and hardly enlightening.) As a result, 
the perturbative solution described by Eq.~\eqref{mh_initialconditions} has one, rather than two, free parameters. 

Setting that parameter (say $\delta_{0,{\rm h}}$) to zero yields the trivial solution $\delta(r)=\chi(r)=\psi(r)=\Sigma(r)=0$. If instead $\delta_0\neq0$, 
homogeneity allows rescaling it to $\delta_{0,{\rm h}}=1$, i.e. the near-horizon solution has no free parameter that can be tuned to ensure that the
solution reduces to a khronon moving with speed $-v$ on flat space at spatial infinity. Indeed, we have verified that the solution obtained by imposing
regularity at the matter/spin-0 horizon and integrating outwards is not asymptotically flat.

\begin{figure}[thbp]
\centering
\includegraphics[width=0.495\textwidth]{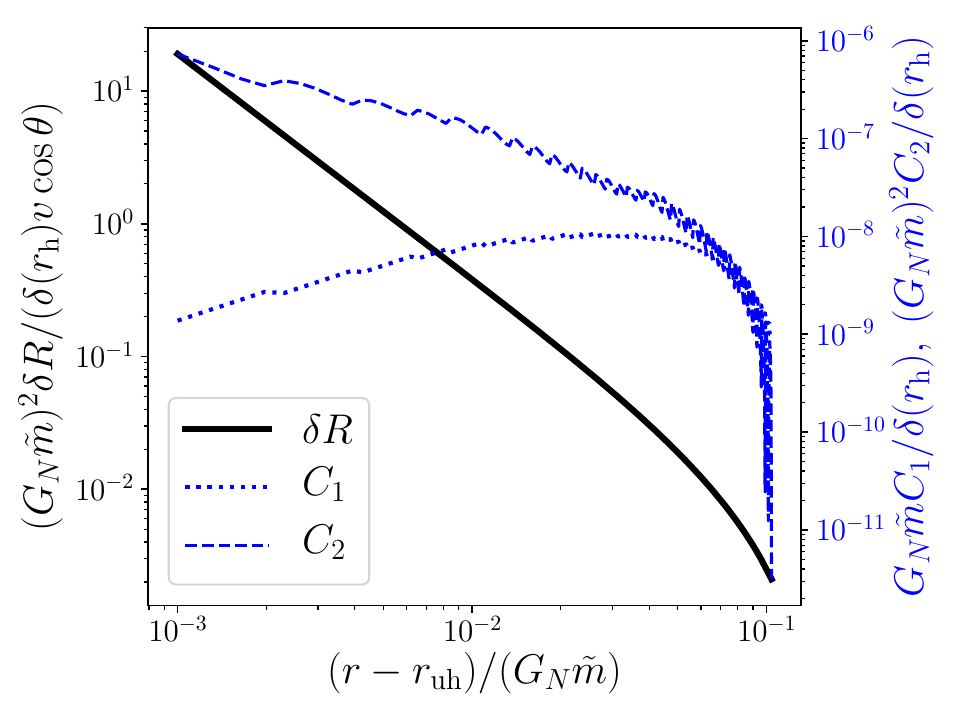}
\caption{${\cal O}(v)$ contribution to the Ricci scalar (left axis) and constraint violations (right axis) near the universal horizon, 
as a function of distance from the latter. The results are in the spin-0 frame,
for the solution
regular at both the matter and spin-0 horizons. 
The theory's parameters are $\a=0.02$, $\b=0.01$ and $\l=0.1$,
corresponding to $\a'=0.02$, $\b'=-4.161$ and $\l'=4.735$.  
Note that this solution is not asymptotically flat, as discussed in the text, and that it is determined
up to a global rescaling. Because of this, we normalize the ${\cal O}(v)$ contribution to the Ricci scalar
by the value of $\delta$ at the matter/spin-0 horizon $r_h$.} \label{uh divergence}
\end{figure}

Moreover, as mentioned in Sec.~\ref{ss:Boundary conditions}, the field equations for the potentials $\delta$, $\chi$, $\psi$ and $\Sigma$ also present a singularity at the universal horizon. Therefore, even if one is willing to accept as physically relevant a BH with non-flat asymptotic boundary conditions, such a solution has no free parameters to tune to impose regularity at the universal horizon either. Indeed, we have verified that integrating the solution inwards from the (regular) spin-0/matter horizon, 
the curvature invariants blow up at the universal horizon (c.f. Fig.~\ref{uh divergence}, 
where we also show the violations of the constraints). 

To further validate this result, we have also tried to first impose regularity at the universal horizon, and then
integrate outwards trying to match with the solution obtained by imposing regularity at the spin-0/matter horizon. In practice, 
we impose regularity at the universal horizon by solving the field equations perturbatively with the ansatz given by Eq.~\eqref{mh_initialconditions}, where $r_h$ is now meant to denote the universal horizon. (Barring
cancellations, analyticity of the potentials is once again required to ensure that
the \ae ther, the two Killing vectors and the geometry are generically regular, i.e. that invariants constructed 
with the curvature tensors, the \ae ther and the Killing vectors remain finite.)

 Rescaling the solution by exploiting again the homogeneity of the problem, we are left with just two 
 free parameters in the perturbative solution near the universal horizon~\footnote{More precisely, 
 if we assume $\Sigma_{0,{\rm h}}\neq0$, we can use the rescaling freedom to set $\Sigma_{0,{\rm h}}=1$. This results 
 in two free parameters, say $\delta_{0,{\rm h}}$ and $\chi_{0,{\rm h}}$. If instead $\Sigma_{0,{\rm h}}=0$, one is still left 
 with two free parameters, say $\delta_{0,{\rm h}}$ and $\chi_{0,{\rm h}}$, and we can then
use the rescaling freedom to set either to 1. Therefore, if $\Sigma_{0,{\rm h}}=0$ one has just one free parameter, 
which makes the matching to the solution regular at the spin-0/matter horizon even more difficult to achieve. 
We have indeed verified that the matching is not possible if one assumes $\Sigma_{0,{\rm h}}=0$.}, 
which we try to tune by matching to the solution that is regular at the spin-0/matter horizon. The latter solution being completely determined, up to a global rescaling, necessary conditions for matching include the continuity conditions
\begin{equation}\label{continuity}
\Delta\left(\frac{\delta'}{\delta}\right)=\Delta\left(\frac{\chi'}{\chi}\right)=\Delta\left(\frac{\psi'}{\psi}\right)=0\,,
\end{equation}
where $\Delta(X'/X)$ denotes the difference between $X'/X$ (with $X=\delta,\chi,\psi$), as
given by the two solutions,  at some matching point between the spin-0/matter horizon and the universal horizon. Note that it does not
make sense to impose continuity of the two solutions ($\Delta X=0$), since we have used the rescaling freedom of the problem to renormalize both (with
a priori different factors). That rescaling
clearly cancels out when considering the ratios $X'/X$. Note also that it does not make sense to
impose continuity of $\Sigma'/\Sigma$, since $\Sigma$ satisfies a first order equation [c.f. Eq.~\eqref{potential dynamical equations Sigma}]. 

Quite unsurprisingly, we have verified numerically that for generic values of the coupling constants, the three conditions of Eq.~\eqref{continuity}
cannot be all satisfied by tuning the two free parameters of the solution regular at the universal horizon. The conclusion is therefore
that even if one gives up asymptotic flatness, for generic values of the coupling constants the universal horizon is 
a finite-area curvature singularity. This is again reminiscent of the occurrence of similar finite-area singularities 
at all but the outermost spin-1 horizon of slowly rotating BHs in Einstein-\ae ther theory~\cite{Barausse:2015frm}. Quite suggestively,
Ref.~\cite{Blas:2011ni} also found that the universal horizon is unstable at second order in perturbation theory
and in the eikonal limit in khronometric gravity, and conjectures that it may give rise to a finite-area curvature singularity.
This instability may be related~\cite{PhysRevD.41.1796} to the universal horizon being also a Cauchy horizon~\cite{Bhattacharyya:2015uxt}. 
While our result is obtained in a completely different framework, it is
interesting that it hints at the same conclusions.

%------------------------------------------------------------------------------------------------------------------------------------------------------------------------------------------------
%------------------------------------------------------------------------------------------------------------------------------------------------------------------------------------------------
\section{\label{sec:level6}The $\a=\b=0$ case}

The results of Sec.~\ref{sec:level5} on the non-existence of slowly moving BH solutions regular everywhere outside $r=0$ apply for generic values of
the coupling constants, i.e. $\a,\b,\l\neq0$. As we have shown, it is possible to attain regularity of the spin-0 and matter horizons (even though at the cost
of giving up asymptotic flatness), but imposing also regularity of the universal horizon remains impossible.

\begin{figure}[thbp]
\centering
\includegraphics[width=0.495\textwidth]{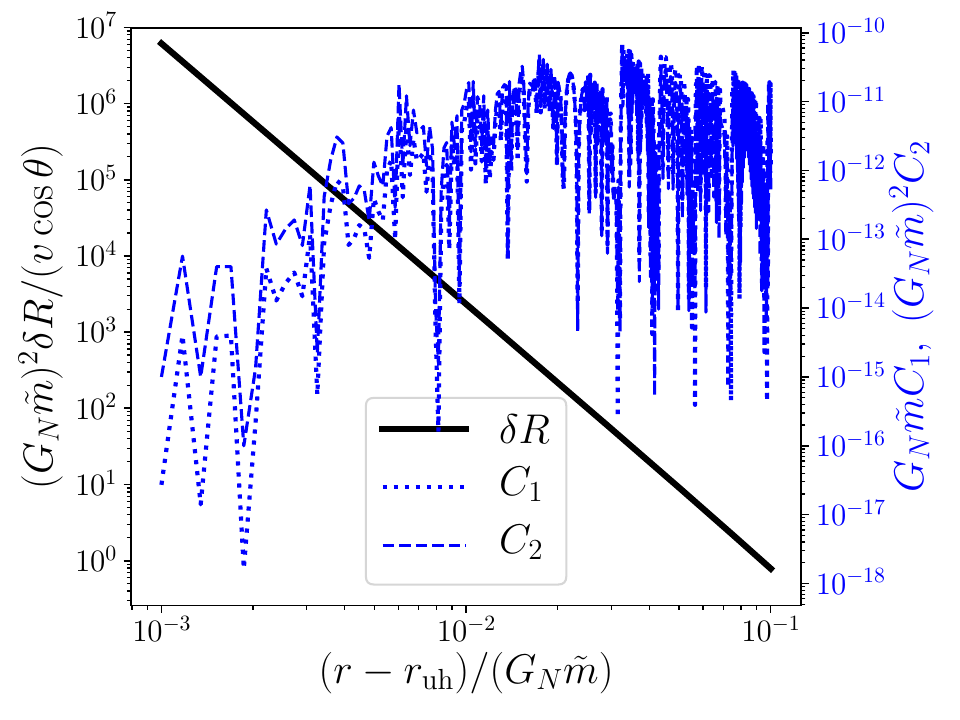}
\caption{${\cal O}(v)$ contribution to the Ricci scalar (left axis) and constraint violations (right axis) near the universal/spin-0 horizon, 
as a function of distance from the latter. The results are for the asymptotically flat solution regular at the matter horizon, and for $\a=0$, $\b=0.01$ and $\l=0.1$.}
\label{uh divergence alpha=0}
\end{figure}

However, if the coupling constants are such that the spin-0 speed $c_0$ diverges, the spin-0 horizon coincides with 
the universal horizon (since the latter is the horizon for modes of infinite speed). Therefore, imposing regularity of the universal, spin-0 and 
matter horizons may become possible in that limit.
From Eq.~\eqref{spin0speed} it follows that $c_0\to\infty$ when $\a\to0$. This limit is particularly attractive as experimentally we have  $|\a|\lesssim 10^{-7}$  
(c.f.~Sec.~\ref{ssec:Experimental constraints}).
Assuming $\a=0$ alone, however, does not avoid the appearance of finite-area singularities at the universal/spin-0 horizon, as can be seen from
Fig.~\ref{uh divergence alpha=0}, 
where we show the divergence of the curvature invariants of the asymptotically flat solution regular at the matter horizon. 

However,
from the experimental limits presented in Sec.~\ref{ssec:Experimental constraints}, it follows that $|\b|\lesssim 10^{-15}$, so 
it is tempting to also set $\b=0$ exactly. Indeed, spherical BH
solutions for $\a=\b=0$ are very simple and known analytically in this limit, and are given by Eqs.~\eqref{fAB_alpha0_a}--\eqref{fAB_alpha0_c}. Note in particular
that the metric matches the Schwarzschild solution in this limit.

By solving the evolution and constraint equations near the metric horizon $r_h$ by imposing regularity there, i.e. with the ansatz of Eq.~\eqref{mh_initialconditions}, 
one immediately finds that $\Sigma$ and $\psi$ must be exactly zero near $r_h$, i.e. $\Sigma={\cal O}(r-r_h)^{n_{\max}}$ and $\psi={\cal O}(r-r_h)^{n_{\max}}$,
where $n_{\max}$ is the order at which the series  of Eq.~\eqref{mh_initialconditions} is truncated. We have indeed verified this for $n_{\max}$ as large as 10 or more.
One reaches the same conclusions by considering series-expanded solutions to the field equations around any other radius (different from the metric horizon). 
Moreover, to further verify that $\Sigma$ and $\psi$ vanish, 
we have replaced $\Sigma(r)=\psi(r)=0$ into the field equations \eqref{potential dynamical equations delta}--\eqref{potential dynamical equations Sigma}. 
The system is in principle overdetermined, but it turns out to consist of just two independent equations:
\bea
\label{beta0eom1}
&\d'(r)+\displaystyle\frac{4(8r^4+4G_N\tilde{m} r^3-27(G_N\tilde{m})^4)}{16r^5-32 G_N\tilde{m} r^4+27(G_N\tilde{m})^4 r} \d(r)\qquad\nn \\
\qquad&-\displaystyle \frac{32r^2}{16r^4-32 G_N\tilde{m} r^3+27(G_N\tilde{m})^4 } \c(r)=0\,,\\
\label{beta0eom2}
&\c'(r)-\d(r)=0\,.\qquad \qquad \qquad \qquad \qquad \qquad \qquad
\eea
Note that these equations do not depend on $\l$, which we have anyway kept different from zero.

Eliminating $\chi$ from Eqs.~\eqref{beta0eom1}--\eqref{beta0eom2} then yields 
\be\label{beta0eom1bis}
\begin{split}
&\left(\frac{r^2}{2}-G_N\tilde{m} r +\frac{27(G_N\tilde{m})^4}{32 r^2}\right) \d''(r)\\ &+\left(2r-\frac{3G_N\tilde{m}}{2}-\frac{81(G_N\tilde{m})^4}{16r^3}\right)\d'(r)
\\ &+\frac{81(G_N\tilde{m})^4}{8r^4}\d(r)=0\,.
\end{split}
\ee
Solving this equation near spatial infinity gives
\be\label{lambda asymptotic delta}
\d(r)=\d_0+\frac{\d_3}{r^3}+\mO\left(\frac{1}{r^4}\right)\,,
\ee
where $\d_0$ and $\d_3$ are integration constants.
This in turn implies, through Eq.~(\ref{beta0eom1}), that $\c(r)$ behaves asymptotically as
\be
\c(r)=\d_0 r+ \c_0 - \frac{\d_3}{2 r^2}+\mO\left(\frac{1}{r^4}\right)\,,
\ee
where $\c_0=\displaystyle-{G_N \tilde{m}}\d_0/{2}$. Replacing this relation in Eq.~\eqref{sensitivity}
and evaluating for $\a=\b=0$ gives a vanishing sensitivity $\sigma=0$.
This result had to be expected from the fact that Eqs.~\eqref{beta0eom1}--\eqref{beta0eom1bis} do not depend on $\l$, 
and that $\sigma$ must go to zero in the general-relativistic limit $\l\to0$.

Moreover, one can push the argument even further, and note that since it is independent of $\l$ and because it must
reduce to the Schwarzschild solution in the general-relativistic limit $\l\to0$, 
the solution to Eqs.~\eqref{beta0eom1} and \eqref{beta0eom2} must simply be the Schwarzschild metric in a weird gauge. 
Indeed, it is easy to check that the metric of Eq.~(\ref{metric_Ansatz}), with $\psi=\Sigma=0$, becomes
the Schwarzschild metric in Eddington-Finkelstein coordinates if one performs the gauge transformation
$\vv'=\vv+v\;\c(r)\,\cos\th+{\cal O}(v)^2$ 
[note that we also need to use Eq.~\eqref{beta0eom2} to set $\c'(r)=\d(r)$].

In spite of this, the khronon field profile is non-trivial [even though its stress energy must vanish through order ${\cal O}(v)$
to allow for the metric to coincide with the Schwarzschild solution, i.e. the khronon is a ``stealth'' field]. In more detail, even though it is
clear that the universal horizon must be a regular surface (since the Schwarzschild metric has no curvature singularity
at $r\neq0$), it is interesting to look for an approximate solution to Eq.~\eqref{beta0eom1bis} near the universal horizon position $\ruh=3 G_N \tilde{m}/2$, at which
Eq.~(\ref{beta0eom1bis}) is singular (because the coefficients multiplying $\delta''$ and $\delta'$ vanish
 at the universal horizon).
For $r\approx \ruh$, Eq.~(\ref{beta0eom1bis}) becomes 
\be
x^2\, \d''(x)+5x\,\d'(x)+2\,\d(x)\approx0\,,
\ee
with $x=r-\ruh$, which yields the general solution
\be \label{hard-soft modes}
\d(x)\simeq C_{\rm h}\, x^{-\sqrt{2}(1+\sqrt{2})}+C_{\rm s}\, x^{\sqrt{2}(1-\sqrt{2})}\,.
\ee
where $C_{\rm h}$ and $C_{\rm s}$ are integration constants (we refer to the mode with coefficient
$C_{\rm h}$ as the ``hard mode'', because it diverges faster than the ``soft mode'' with coefficient  $C_{\rm s}$).

While both the soft and hard modes diverge as $r\to\ruh$, it is easy to check that the curvature invariants $R$, $R_{\a\b}R^{\a\b}$
and $R_{\a\b\gamma\delta}R^{\a\b\gamma\delta}$ are regular (which must be the case since the metric is Schwarzschild in disguise). 
One can look, however, also at curvature invariants constructed with the \ae ther vector and with the Killing vectors $\partial_\vv$
and $\partial_\phi$. The only non-trivial invariant [at order ${\cal O}(v)$] among
these is 
\be
R_{\m\n\a\b}\, u^\m\, u^\a\, (\p_\vv)^\nu\,(\p_\vv)^\b \propto  \cos\th \;x^3\; \d(x)\,.
\ee
Using Eq.~\eqref{hard-soft modes}, this becomes
\be
R_{\m\n\a\b}\, u^\m\, u^\a\, (\p_\vv)^\nu\,(\p_\vv)^\b \propto  \cos\th \left(C_{\rm h} \,x^{n_h} +C_{\rm s}\, x^{n_s} \right)\,, \\
\ee
with $n_h={1-\sqrt{2}}<0$ and $n_s={1+\sqrt{2}}>0$. Therefore, the hard mode produces a singularity at the universal horizon,
while the soft mode is physically well-behaved.

\begin{figure}[thbp]
\centering
\includegraphics[width=0.495\textwidth]{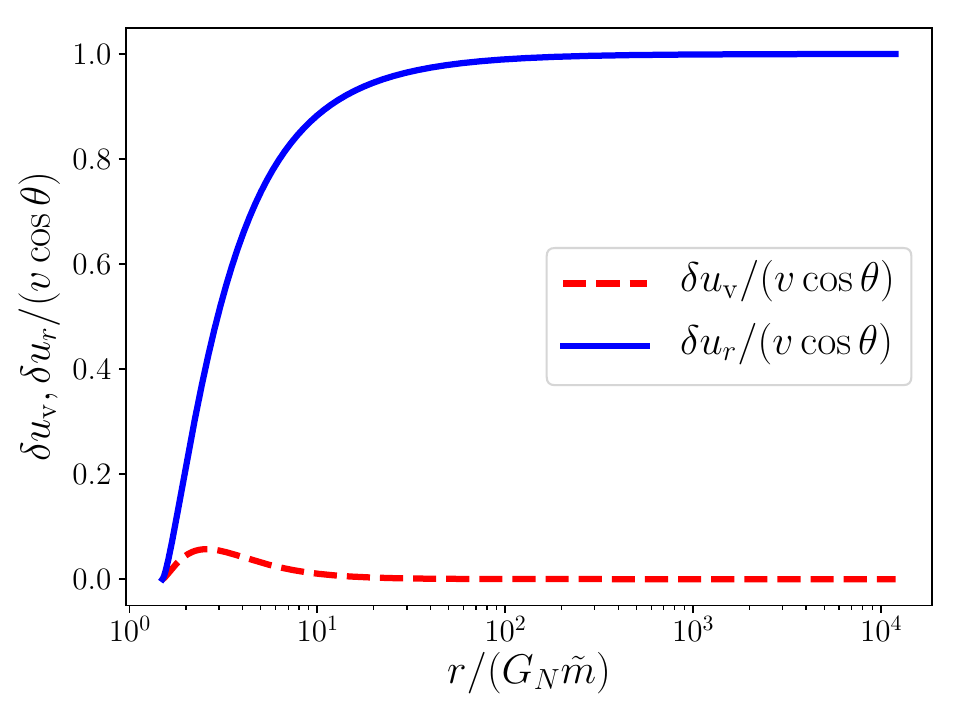}
\caption{${\cal O}(v)$ \ae ther perturbations $\d u_\m$ for the unique regular solution of the $\a=\b=0$ case, outside the universal/spin-0 horizon.}
\label{deltauv_deltaur}
\end{figure}

One can therefore set $C_{\rm h}=0$ in Eq.~\eqref{hard-soft modes}, choose $C_{\rm s}=1$ by rescaling the solution (without loss of generality),
and then use Eq.~\eqref{hard-soft modes} to provide initial conditions at $r=\ruh(1+\epsilon)$ (with $\epsilon\ll1$)
for Eq.~\eqref{beta0eom1bis}. Integrating that equation outwards and matching to Eq.~\eqref{lambda asymptotic delta},
one can then extract the integration constants $\delta_0$ and $\delta_3$. (No shooting procedure is needed in this case
as the initial conditions at $r=\ruh(1+\epsilon)$ are completely determined.) Finally, one can rescale the obtained solution
by a global factor to impose the boundary condition $\delta_0=-1$ (c.f. Sec.~\ref{ss:Construction of the Ansatz}). Eq.~\eqref{beta0eom1} then allows one
to obtain $\chi$. The resulting solution for 
\begin{gather}
\frac{\delta u_\vv}{v \cos\theta}=-\frac{\left(1-A(r)^2 f(r)\right) \left(1+A(r)^2 f(r)\right)^2}{8 A(r)^3 B(r)}\delta(r)\,,\label{solAE1}\\
\frac{\delta u_r}{v \cos\theta}=\frac{1-A(r)^4 f(r)^2}{4 A(r)}\delta (r)\,,\label{solAE2}
\end{gather}
is shown in Fig.~\ref{deltauv_deltaur}. 
Note that both quantities are regular at the universal horizon [as can also be verified analytically
using the soft mode of the solution given by Eq.~\eqref{hard-soft modes} into Eqs.~\eqref{solAE1} and \eqref{solAE2}], which confirms that the khronon field is regular there. Also note that the \ae ther field is not trivial (e.g. it is not static), even if one transforms it into the gauge where 
the metric becomes Schwarzschild in Eddington-Finkelstein coordinates through linear order in the velocity $v$.
%------------------------------------------------------------------------------------------------------------------------------------------------------------------------------------------------
%------------------------------------------------------------------------------------------------------------------------------------------------------------------------------------------------
\section{\label{sec:level7}Conclusions}
We have studied non-spinning BHs moving slowly relative to the preferred foliation of khronometric theory (the low energy limit of Ho\v rava gravity).
We have done so by reducing the field equations (through first order ${\cal O}(v)$ in the velocity relative to the preferred frame) to a system of ordinary differential equations in the radial coordinate, thanks to suitable
ans{\"a}tze for the metric and khronon fields, inspired by the cylindrical symmetry of the system.
We have solved these equations numerically, trying to impose both asymptotic flatness and regularity at the multiple BH horizons that
exist in Ho\v rava gravity, i.e. the matter horizon; the horizons for spin-0 and spin-2 gravitons, and the universal horizon for modes whose speed diverges in the UV.
While regularity at the spin-2 horizon does not pose any particular issue (as expected), regularity at the other horizons is more problematic. 

We have indeed found that if one imposes regularity at the matter horizon and asymptotic flatness, slowly moving BHs necessarily present (for generic values of the dimensionless coupling parameters $\a$, $\b$ and $\l$) 
a curvature singularity at the spin-0 horizon (which lies inside the matter horizon for experimentally viable values of $\a$, $\b$ and $\l$). 
By waiving the requirement of asymptotic flatness, solutions that are regular at the matter and spin-0 horizons can be obtained, 
but are singular further inside, as they develop a curvature singularity at the universal horizon. These pathological features cast doubts on 
the viability of the theory for generic values of the coupling parameters, although these curvature singularities
(strictly speaking) simply signal  that our slow-motion approximation (which assumes implicitly that the ``potentials'' are small) breaks down.
Also, these curvature singularities will probably be smoothed out~\cite{Blas:2014aca} by the higher energy UV corrections $L_4$ and $L_6$ in the Ho\v rava gravity action [c.f.~Eq.~\eqref{action HL}].

Nevertheless, adopting generic values of the coupling parameters $\a$ and $\b$ is not necessarily justified. 
The experimental constraints that we have reviewed in Sec.~\ref{ssec:Experimental constraints} imply  $|\a|\lesssim 10^{-7}$ and $|\b|\lesssim 10^{-15}$, 
hence it would be quite natural to assume that  $\a$ and $\b$ are {\it exactly} zero. In that case, slowly moving BH solutions exist and are regular everywhere outside the central $r=0$ singularity. 
More importantly, even though the gravitational theory is different than GR (because $\l\neq0$), the khronon is a non-trivial ``stealth'' field in these regular BH solutions, 
whose metric therefore reduces exactly to the Schwarzschild one. This implies in particular that BH sensitivities are exactly zero for $\a=\b=0$,  
hence BH binaries do not emit dipolar radiation in this limit, nor do they deviate 
from GR at Newtonian order in the conservative sector [c.f.~Eq.~\eqref{eq:New_active}].\footnote{Note that from Eq.~\eqref{Pdot-AE} it follows immediately
that dipolar radiation (regulated by the coefficient ${\cal C}$)
and all other scalar effects [i.e. the terms depending on the spin-0 velocity in Eq.~\eqref{Pdot-AE}]  
vanish automatically in the case when $\a=\b=0$ (even if the sensitivities were non-zero), 
because the spin-0 speed diverges. That means that the spin-0 mode becomes non-dynamical, and in particular that it does not produce a GW flux.} 
Indeed, these results confirm the conclusion of Ref.~\cite{2014arXiv1407.1259L}, namely that vacuum asymptotically flat solutions to 
khronometric theories with $\a=\b=0$ coincide with the general
relativistic ones even though $\l\neq0$. 

We therefore expect GW generation to agree exactly with GR even at higher PN orders (quadrupolar emission and higher)
if $\a=\b=0$. This is quite important from an observational point of view, because it implies
that even if our results for the appearance of finite-area singularities in moving BHs were just an artifact of
the breakdown of our approximation scheme, and moving BHs turned out to be regular (away from $r=0$),
deviations from GR in GW generation are bound to be small. Indeed,
in such a situation,  deviations away from the GR predictions for GW emission should be expected to be of (fractional) order ${\cal O}[\max(\alpha,\beta)]\sim 10^{-7}$
 for viable values of $\alpha,\beta\neq0$. Such small deviations are unlikely to be observable with present and future GW detectors~\cite{Barausse:2016eii},
although the viable parameter space for $\a,\b$ may further be shrunk by observations of GW and electromagnetic-wave propagation in multimessenger events.

However, if finite-area singularities do indeed form in moving BHs (though perhaps smoothed out by UV corrections~\cite{Blas:2014aca}), they could produce firewall-like~\cite{Almheiri:2012rt} surfaces 
that may in principle be tested with GW echoes~\cite{Barausse:2014pra,Barausse:2014tra,Cardoso:2016rao} or stochastic background measurements from LIGO/Virgo~\cite{Barausse:2018vdb}. As for $\l$, it is likely that improved 
constraints on it may
come from cosmology. As mentioned, Ref.~\cite{Afshordi:2009tt} showed that CMB measurements constrain $0\neq \l\lesssim 10^{-2}$ when $\a=\b=0$, and one would expect this
bound to be robust when small but finite values of $\a$ and $\b$ are considered. Further improvements may come from future CMB experiments and/or galaxy surveys.

\begin{acknowledgments}
We would like to warmly thank Daniele Steer, Niayesh Afshordi and Eugene Lim for providing insightful comments about this work.
We also thank Ted Jacobson, Thomas Sotiriou and Diego Blas for useful discussions 
about Lorentz violating gravity. 
O.R. acknowledges support from a Lagrange Thesis Fellowship of the Institut Lagrange de Paris
(ILP LABEX ANR-10-LABX-63), supported through the Investissements
d'Avenir program under reference ANR-11-IDEX-0004-02.
This project has received funding from the European Union's Horizon 2020 research and innovation program under the
Marie Sklodowska-Curie grant agreement No 690904.
\end{acknowledgments}

\bibliography{shortbib}

\end{document}